\documentclass[%
 reprint,
superscriptaddress,
preprintnumbers,
nofootinbib,
 amsmath,amssymb,
 aps,
 prx,
floatfix,
]{revtex4-2}

\usepackage{amssymb} 
\usepackage{graphicx}
\usepackage{dcolumn}
\usepackage{bm}
\usepackage[colorlinks,
           linkcolor=red,
           citecolor=blue,
           allcolors=blue
            ]{hyperref}
\usepackage{cancel}
\usepackage{xcolor}
\usepackage{graphicx}
\usepackage{dcolumn}
\usepackage{bm}
\usepackage{bbm, dsfont}
\usepackage{ulem}

\makeatletter

\newcommand{\Rmnum}[1]{\expandafter\@slowromancap\romannumeral #1@}
\makeatother
\usepackage{amsthm,amsmath,amssymb}
\newcommand{\gn}{g_{\rm aNN}}
\newcommand{\dm}{\delta m^2}

\renewcommand{\eqref}[1]{Eq.(\ref{#1})}

\begin{document}
\preprint{KEK-QUP-2026-0001}
\title{\textbf{Earth Matter Enhanced Axion Dark Matter Search}}

\author{Xiaofei Huang}
\email[]{These authors contributed equally to this work}
\affiliation{School of Instrumentation and Optoelectronic Engineering, Beihang University, Beijing 100191, China}
\author{Xiaolin Ma}
\email[]{These authors contributed equally to this work}
\affiliation{International Center for Quantum-field Measurement Systems for Studies of the Universe and Particles (QUP, WPI), High Energy Accelerator Research Organization (KEK), Oho 1-1, Tsukuba, Ibaraki 305-0801, Japan}
\author{Zitong Xu}
\email[]{Contact author: zitong.xu@ntu.edu.sg}
\affiliation{Quantum Science and Engineering Centre, Nanyang Technological University, Singapore, 639798, Singapore}
\affiliation{School of Physical and Mathematical Sciences, Nanyang Technological University, Singapore, 639798, Singapore}

\author{Itay M. Bloch}
\affiliation{CERN, Theoretical Physics Department, Geneva, Switzerland}
\affiliation{Physics Department, Technion - Israel Institute of Technology, Haifa 3200003, Israel}

\author{Kai Wei}
\email[]{Contact author: weikai@buaa.edu.cn}
\affiliation{School of Instrumentation and Optoelectronic Engineering, Beihang University, Beijing 100191, China}
\affiliation{Quantum Science and Technology College, Beihang University, Beijing 100191, China}
\affiliation{Hefei National Laboratory, Hefei, 230088, China}

\date{\today}

\begin{abstract}
{Laboratory searches for ultralight axion dark matter (DM) have traditionally assumed the terrestrial density of axions is equal to the average density of DM in the solar system. However, quadratic couplings to matter introduce a non-trivial field profile near the Earth.
In this work, we present the first dedicated experimental implementation of an environment-aware axion DM wind search framework. 
Leveraging the extreme sensitivity of a K--Rb--$^{21}$Ne comagnetometer to pseudo-magnetic fields induced by axion DM, we analyzed our data in the context of the massively enhanced local gradient of axions due to interactions with matter, though no signal candidates were found. Consequently, we have set the most stringent limits on axion-neutron derivative interactions for masses
$m_a \in [0.041, ~28.9]~\rm feV$,
improving from previous experiments that ignore terrestrial matter effects by as much as three orders of magnitude for certain masses.
Our work highlights the necessity of accounting for environmental modifications in precision frontier experiments and demonstrates how geophysical variations can be harnessed to act as a natural amplifier for DM possibly enabling future detection in parts of the parameter space that were previously beyond reach.}

\end{abstract}

\maketitle

\section{Introduction}

The nature of Dark Matter (DM) is an unsolved puzzle, and the search for DM remains one of the most promising endeavors in fundamental physics. Axions stand out among various DM candidates, originally postulated as a solution to the strong CP problem~\cite{Peccei:1977hh, Weinberg:1977ma, Wilczek:1977pj}, a variety of production mechanisms exist which produce them at the correct abundance to explain the observed DM density~\cite{duffy2009axions,Marsh_2016,Preskill_1982,Abbott_1982,Dine_1982,DiLuzio:2021gos}.

While the axion mass is unknown, with an uncertainty spanning many orders of magnitude, a significant experimental effort has targeted the particularly interesting ultralight regime ($m_ac^2 \ll 1\,\rm{eV}$)~\cite{Hu:2000ke,Cheong:2024ose}, in which the axion DM behaves not as individual particles but as a coherent classical field, oscillating with a frequency determined by its mass. 
Over the past decades, a vast array of terrestrial experiments has been deployed to detect this field~\cite{DiLuzio:2020wdo, Adams:2022pbo, OHare:2024nmr}. 
However, these searches typically assume the local axion field amplitude is unaffected by the Earth, taking the results of galactic simulations that assume negligible DM-matter interactions\footnote{Recently there are some papers also considering the non-trivial dark matter profile around Earth surface, for example, refs.~\cite{Wilson:2025lhq,Arza:2025kuh}.}.

While this assumption is valid for canonical quantum chromodynamics (QCD) axions models~\cite{Kim:1979if,Shifman:1979if,Dine:1981rt,Zhitnitsky:1980tq}, it fails to capture the rich phenomenology arising for bosons with larger non-derivative interactions with the Standard Model~\cite{Stadnik:2015kia,Hees:2018fpg}. One such example is ``light QCD axions''~\cite{Hook:2018jle,DiLuzio:2021pxd}, which solve the strong CP problem and feature an enhanced coupling to gluons. This coupling induces an in-medium potential that shifts the axions' effective mass in matter. In dense objects such as neutron stars~\cite{Hook:2017psm} and white dwarfs~\cite{Balkin:2022qer} this potential is known to lead to observable effects. However, only recently~\cite{Banerjee:2025dlo,delCastillo:2025rbr} was it realized that for light axion DM, the Earth's matter can also have important effects, prompting a reassessment of how terrestrial searches should be interpreted. In particular, for a wide range of parameter space, the axion DM field near the Earth's surface can be suppressed while its gradient, especially in the radial direction, can be drastically enhanced~\cite{Banerjee:2025dlo,delCastillo:2025rbr}.

This terrestrial modification of the DM wind fundamentally reshapes the experimental landscape. Detectors targeting axion-photon couplings (e.g.~cavities) tend to rely on a direct coupling to the axion amplitude, which may have significantly reduced sensitivity in regions where the field is screened by Earth's matter. 
Conversely, the enhanced gradient offers a novel and amplified signature for experiments hunting for the axion wind, \textit{e.g.}, spin-based sensors~\cite{jiang2021search,Wu:2019exd,CASPERSLIDES,wei2025dark,xu2024constraining,Huang2025,Bloch:2022kjm,Bloch:2021vnn,Lee:2022vvb,Gavilan-Martin:2024nlo,Bloch:2019lcy,Alonso:2018dxy}, which search for interactions that couple to the spatial derivative of the DM field.
Despite the profound implications of this effect, no experimental search has yet been optimized to exploit this gradient enhancement.

In this paper, we report the first dedicated search for this type of axion DM  that explicitly accounts for and leverages the Earth's matter effect. 
Using an ultrahigh-sensitivity self-compensation alkali-noble-gas comagnetometer with the sensitive axis tuned to align with Earth radial direction, our experimental design is uniquely suited to detect the enhanced signal that would otherwise escape conventional strategies. 
With no DM signal being observed, we establish the most stringent laboratory constraints on axion-neutron coupling to date  
for a vast region of the mass-decay constant parameter space.
In the mass range where the Earth's environmental effect is most pronounced, our limits surpass those derived from astrophysical bounds and other terrestrial experiments relying solely on the linear coupling by more than two orders of magnitude, demonstrating the importance of the interplay between different DM-matter couplings.

\begin{figure*}
    \centering
\includegraphics[width=0.9\linewidth]{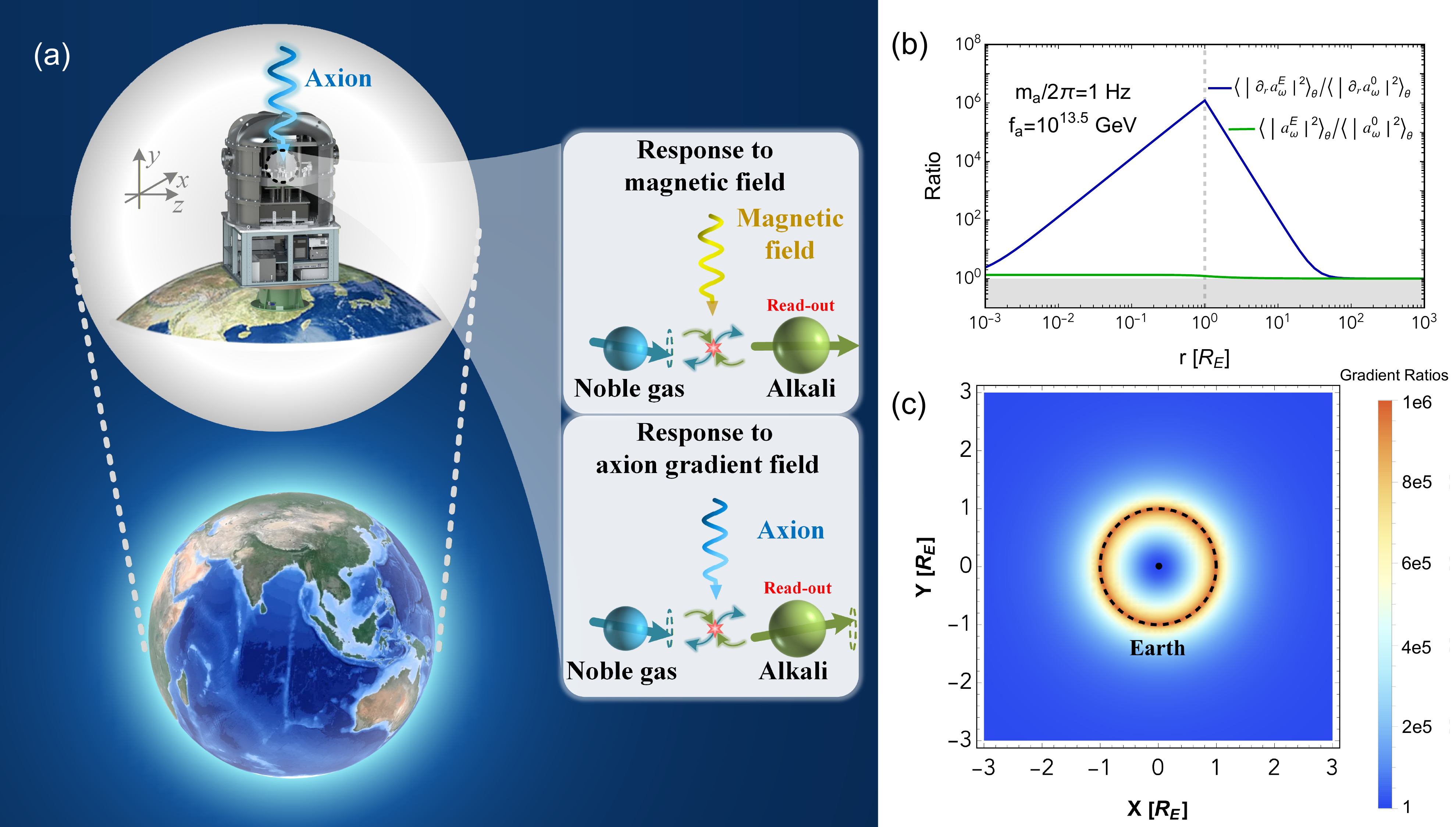}
    \caption{The Earth-effected axion DM and its detection principle. (a)
    Due to the quadratic coupling of axions and fermions, Earth's dense matter changes the effective mass of the axion inside it.
 For a generic quadratic coupling, the effect can be analogous to light traveling between two media with vastly different refractive indices. The presence of the Earth introduces a new small length scale with its radius, and at the surface of the Earth, the radial direction gradient of the axion field will be greatly enhanced, providing an excellent opportunity for a highly sensitive comagnetometer whose sensitive axis is pointing along the radial direction.
    Operating in the self-compensation regime, our coupled alkali-noble-gas comagnetometer is specifically configured with the sensitive axis perpendicular to the ground to benefit from the enhancement.
    For transverse magnetic-field noise, the noble-gas nuclear spins adiabatically follow the perturbation and generate an opposing effective field: it largely cancels the noise experienced by the alkali-metal spins, leaving their orientation nearly unchanged. In contrast, an axion-gradient-induced anomalous field couples primarily to the nuclear spins, driving transverse precession that is read out by the alkali-metal spins acting as an in-situ ultrahigh-sensitivity magnetometer.
    (b) Radial profiles of the axion field amplitude (blue) and gradient (green), normalized to their unenhanced values with angular dependence averaged. We choose a Compton frequency $m_a/2\pi = 1~\mathrm{Hz}$ and a decay constant $f_a = 10^{13.5}~\mathrm{GeV}$ (corresponding to the regime $k R_{\mathrm{E}} < q R_{\mathrm{E}} \lesssim \hbar$ for demonstration. The enhancement of the radial gradient at the surface scales approximately as $1/(k R_{\mathrm{E}})\times \left((f_a)^c/f_a\right)^2$(see the definition of $(f_a)^c$ in main text.), while the field amplitude remains nearly unchanged. 
    (c) Spatial distribution of the axion gradient enhancement ratio for the same parameter. A pronounced enhancement of the gradient is visible near the Earth’s surface, while the effect rapidly attenuates both inside the Earth and far outside it.
        }
\label{fig:illu}
\end{figure*}

\section{Principle}

The primary objective of this experiment is to search for axion DM interacting with nuclear spins via the gradient coupling as shown in Fig.~\ref{fig:illu}(a). In the non-relativistic regime, this interaction is described
by the Hamiltonian: 
\begin{align}
\mathcal{H}=2g_{\mathrm{aNN}}\bm \nabla a\cdot \bm I_{\rm N}
\end{align}
where $\bm I_{\rm N}$ is a nuclear spin, which could include contributions from neutron and proton angular momenta, $\bm \nabla a$ is the gradient of the axion field, which is a stochastic field oscillating with a typical central frequency $\omega\sim m_a$~\cite{derevianko2018detecting,PhysRevD.103.115004,lisanti2021stochastic,Bloch:2021vnn}. $g_{\mathrm{aNN}}$ represents the effective axion-nucleus coupling, in this paper we only consider the axion-neutron coupling and parametrize $g_{\mathrm{aNN}}\equiv \epsilon_n g_{\rm an},~g_{\rm an}\equiv c_n/f_a $ with $\epsilon_n$ being the neutron spin fraction of the nucleus, $g_{\rm an}$ being the axion-neutron coupling strength, and $c_n$ being a model-dependent and fermion-dependent parameter. The coherently oscillating axion DM induces an anomalous pseudo-magnetic field defined as $\bm b\equiv 2g_{\mathrm{aNN}}\bm \nabla a/\gamma_n$, with $\gamma_n$ being the nuclear gyromagnetic ratio. 

The signal $\bm b$ is governed by the spatial distribution of the local axion DM field, which has often been taken as a simple plane-wave $\propto {\rm sin}({\bf k}_a\cdot  {\bf r}\pm \omega t)$. This could be drastically modified when accounting for the matter-effect of the Earth on the axion DM. 

After the QCD phase transition, the axion-nucleon interaction Lagrangian is given by~\cite{graham2015experimental,DiLuzio:2020wdo}:
\begin{equation}
\label{eq:axion_nucleon_int}
\begin{aligned}
\mathcal{L}&= -\sigma_{\rm N} \bar{N}N \sqrt{1-\beta\sin^2\left(\frac{a}{2f_a}\right)} \\
&\simeq \frac{1}{2}\frac{z}{(1+z)^2}\sigma_{\rm N} \bar{N}N \frac{a^2}{f_a^2}\,,
\end{aligned}
\end{equation}
where $\sigma_{\rm N} \simeq 50$~MeV is the nucleon sigma term, $z=m_u/m_d$ is the up/down quark mass ratio, $\beta=4zm_u/(m_u+m_d)^2$~\cite{Banerjee:2025dlo,Alarcon:2021dlz}, and $f_{a}$ is the decay constant of axion. The second equality follows from a leading-order expansion in $a/f_a$, which is highly accurate for the axion dark matter (DM) background where $a/f_a \ll 1$. 
This shift-symmetry breaking coupling of the QCD axion to gluons induces a correction to the squared axion mass in the presence of a background nuclear number density $n$,
\begin{align}
    \delta m^2\simeq -\frac{z}{(1+z)^2}\frac{1}{f_a^2}\sigma_{\rm N}n,
    \label{eq:massshift}
\end{align}
with $z$ being the up-to-down quark mass ratio.  For light QCD axions~\cite{Hook:2018jle,DiLuzio:2021pxd,DiLuzio:2021gos}, the canonical relation $m_a f_a\simeq m_\pi f_\pi$, (e.g., KSVZ~\cite{Kim:1979if,Shifman:1979if} or DFSZ~\cite{Dine:1981rt,Zhitnitsky:1980tq}, with $f_{\pi}$ being the decay constant of pion) is modified to $m_a f_a\simeq \sqrt{\epsilon} m_\pi f_\pi\ll m_\pi f_\pi$ without requiring a large tuning of parameters. Hence, Earth's substantial nucleon density creates an attractive potential well, which induces a negative shift in the axion's squared mass, $m_{\rm eff}^2 = m_a^2 - \delta m^2$ inside Earth. For small $\epsilon$s, this leads to significant deviations from the naive picture.  

The physical intuition is that, by conservation of energy, an axion entering the potential well must increase its kinetic energy to compensate for the reduced effective mass. Consequently, the axion momentum magnitude inside the Earth, $q$, is larger than its outside momentum, $k$:
\begin{align}
q = \sqrt{k^2 + \delta m^2} \gtrsim  k.
\end{align}
This is analogous to the modification of a photon's wavelength (inversely proportional to its momentum) as it travels from a low refractive index medium to a high refractive index one. Hence, as the axion wave encounters the sharp discontinuity of the Earth's potential, it experiences an impedance mismatch. This mismatch causes the majority of the incoming dark matter amplitude to be reflected at the surface of Earth. Consequently, the field amplitude $a$ on the surface can be screened~\cite{Stadnik:2021qyf,Banerjee:2025dlo}, lowering it below the naive plane wave expectation, while the gradient is increased.

\subsection{Axion DM EOM with Earth-Induced Matter Effects}
\label{sec:axion_earth}
To quantitatively determine the field profile, one has to solve the axion equation of motion in the presence of the Earth. We model the Earth as a perfectly uniform sphere of radius $R_{\rm E}$, which is a valid approximation when $|\delta m R_{\rm E}|\lesssim 1$, corresponding to $f_a\gtrsim 10^{13}$ GeV of the experiment interested region. Within that sphere, we modify the squared axion mass according to Eq.~\ref{eq:massshift}. By imposing conservation of energy, requiring continuous differentiability at Earth's surface, and matching to the standard DM solution far from the Earth, the general solution is obtained~\cite{Banerjee:2025dlo}.

The equation of motion (EOM) for the axion field $a(\mathbf{r},t)$ in the Earth-centric frame is:
\begin{equation}
\label{eq:EOM1}
    \begin{cases}
        (\Box + m_a^2)a(\mathbf{r},t) = 0 \,, & r > R_{\rm E} \\
        (\Box + m_a^2 - \delta m^2)a(\mathbf{r},t) = 0 \,, & r < R_{\rm E}
    \end{cases}\,.
\end{equation}

Decomposing the field into energy modes $a(\mathbf{r},t) \simeq \sum_{\omega} f_\omega a_\omega(\mathbf{r})e^{-i\omega t+\phi_\omega} + \text{h.c.}$, each mode satisfies:
\begin{equation}
\label{eq:EOM2}
    \begin{cases}
        \left(\omega^2 - m_a^2 + \nabla^2\right)a_\omega = 0 \,, & r > R_{\rm E} \\
        \left(\omega^2 - m_a^2 + \nabla^2 +\delta m^2\right)a_\omega = 0 \,, & r < R_{\rm E}
    \end{cases}\,.
\end{equation}

Matching boundary conditions at $r=R_{\rm E}$ requires the continuity of the field and its derivatives. Furthermore, for $r \gg R_{\rm E}$, the solution must asymptotically recover the standard DM wind: \begin{align}
\label{eq:AsympAxionDMSol}
a(\mathbf{r},t)\Big|_{r\gg R} \longrightarrow 2a_0 \cos{\left(\omega_a t - \mathbf{k}\cdot\mathbf{r}\right)}\,,
\end{align}
where $a_0$ is determined by the local dark matter density. Here we  consider the monochromatic axion DM model without stochasticity for derivation simplicity, while recovering the DM stochastic effect in the Supplemental Material~\cite{sm} Sec~\ref{sec:stochastic}. Utilizing the azimuthal symmetry of the scattering problem, we decompose the solution into spherical harmonics with azimuthal number $m=0$:
\begin{align}
\label{eq:oursolout}
    a^{\rm out}_\omega(r,\theta) &= a_0\sum_{\ell=0}^{\infty} (2\ell + 1) i^\ell j_\ell(kr) P_\ell(\cos \theta)\nonumber\\
    &+ \sum_{\ell=0}^{\infty} c_\ell^{\text{out}} h_\ell^{(1)}(kr) P_{\ell}(\cos\theta) \, , \nonumber\\
    a^{\rm in}_\omega(r,\theta) &= \sum_{\ell=0}^{\infty} c_\ell^{\text{in}} j_\ell(qr) P_\ell(\cos \theta)\,,
\end{align}
where $j_\ell$ and $h_\ell^{(1)}$ are the spherical Bessel and Hankel functions of the first kind, respectively. Solving the boundary conditions yields the coefficients~\cite{Banerjee:2025dlo}:
\begin{align*}
\label{eq:al}
a^{\rm out}_{\omega,\ell}(r,\theta) &= a_0\frac{ i^{\ell} (2 \ell+1)P_\ell(\cos\theta) }{Q_\ell} \\\times
&\left[ y_\ell(k r)\text{Im}(Q_\ell) + j_\ell(k r)\text{Re}(Q_\ell) \right] \, , \\
a^{\rm in}_{\omega,\ell}(r,\theta) &= -a_0\frac{i^{\ell} (2 \ell+1)P_\ell(\cos\theta) }{k R^2 Q_\ell} j_\ell(q r)\,,
\end{align*}
with the auxiliary functions defined as:
\begin{align*}
\text{Re}(Q_\ell) &= k j_\ell(q R_{\rm E}) y_{\ell+1}(k R_{\rm E}) - q j_{\ell+1}(q R_{\rm E}) y_\ell(k R_{\rm E})\,,\\
\text{Im}(Q_\ell) &= q j_{\ell+1}(q R_{\rm E}) j_\ell(k R_{\rm E}) - k j_\ell(q R_{\rm E}) j_{\ell+1}(k R_{\rm E})\,.
\end{align*}

While the general solution implies a summation over all $\ell$, the relevant physics at the Earth's surface depends on the parameter $kR_{\rm E}$. For $kR_{\rm E} \ll 1$ (which holds for $f \lesssim 10$~Hz), we can Taylor expand the solution. Leading order contributions scale as $(kR_{\rm E})^\ell$ for generic $q$s, implying that higher-$\ell$ modes are suppressed. Thus, for low-frequency experiments such as comagnetometers, we may focus on the $\ell=0$ and $\ell=1$ modes, while in practice, the monopole mode dominates the signal in our region, we only focus on the $\ell=0$ mode in the following discussion.

A key feature of the matter effect is the modification of the axion gradient. For the $\ell=0$ mode, the radial derivative at the surface is:
\begin{equation}
\partial_r a_{\omega,0}(R_{\rm E}) = \frac{e^{-ikR}a_0}{R_{\rm E}}\frac{qR_{\rm E}\cos(qR_{\rm E})-\sin(qR_{\rm E})}{qR\cos(qR_{\rm E})-ikR\sin(qR_{\rm E})}\,.
\label{eq:da0}
\end{equation}

For $qR_{\rm E} \lesssim 1$, the gradient magnitude is generically enhanced relative to the vacuum expectation $ka_0$ by a factor of $1/(kR_{\rm E})$, or $1/(kR_{\rm E})^2$ when $\cos(qR_{\rm E})=0$ (so-called resonances).
 In general, away from resonances (which only appear when $\delta m R_{\rm E}>1$),  the typical radial gradient is approximately:
\begin{align}
   & \partial_r a^{\rm E}_{\omega}(R_{\rm E}) \nonumber\\
   &\sim k a_0 \left(1+\max  \left[1,\frac{1}{k R_{\rm E}}\right]\,\min \left[1, \left(\frac{(f_a)^c}{f_a}\right)^2 \right]\right) \ ,
    \label{eq:radialGradientEarth}
\end{align}
while the amplitude scales as 
\begin{align}
    a^{\rm E}_{\omega}\sim a_0 ~{\rm min}\left[1, \frac{f_a}{(f_a)^c}\right]
\end{align}
with $a_0=\sqrt{2\rho_{\rm DM}}/m_a$ as the vacuum axion DM amplitude, and  $(f_a)^c(m_a)\equiv 3\times10^{13}~{\rm GeV}/\sqrt{1+k^2R_{\rm E}^2}$ defined as the critical value of the decay constant for preventing the Earth sourcing axion~\cite{Banerjee:2025dlo}.
Crucially, in the regime of interest for our experiments $kR_{\rm E}<qR_{\rm E}<1$, the gradient of the field is enhanced while its amplitude remains unperturbed. 
In particular, the radial gradient at the Earth's surface is enhanced by a factor of around $1/kR_{\rm E}\times\left((f_a)^c/f_a\right)^2$ compared to the axion DM away from the Earth (See Fig.~\ref{fig:illu} and Sec.~\ref{sec:axion_earth} below for details). This amplification opens new sensitivity windows for experiments coupled to the gradient of the axion field. 

\subsection{Experimental Setup and Data Acquisition}

The discussion above establishes the form of the effective fields induced by dark matter interactions. To access these signatures experimentally, a detection scheme capable of converting such weak effective fields into measurable signals is required. In this work, this role is fulfilled by an atomic comagnetometer, whose operating principles are described below.

The atomic comagnetometer uses 3 interacting spin-ensembles and two near-infrared lasers to convert the Earth-enhanced dark matter gradient field into optical readout as shown in Fig.~\ref{fig:ExSe} in the Supplemental Material~\cite{sm}. The K--Rb--$^{21}$Ne ensembles form a coupled spin-exchange system in which the alkali-metal and noble-gas spins are coupled through successive spin-exchange processes. Optically pumped potassium polarizes the Rb atoms through fast spin exchange collisions. On longer timescales, the collisions of Rb atoms with $^{21}$Ne polarize the nuclear spin of the noble gas, creating a long-lived nuclear polarization that evolves on a much slower timescale than alkali ensembles, which are mostly electron-spin based. 
The sensitive axis is specifically oriented in the Earth's radial direction, thereby achieving sensitivity to the matter-enhanced axion DM field.

\begin{figure*}[!htbp]
    \centering
    \includegraphics[width=1\linewidth]{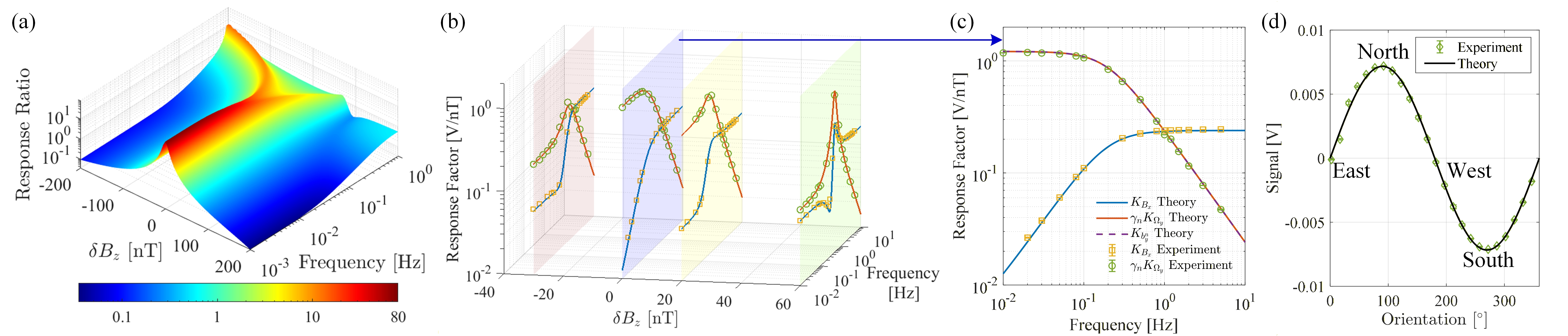}
    \caption{Comparison of responses to classical magnetic fields and non-magnetic inputs. (a) Simulated ratio between the comagnetometer responses to the anomalous field $b_y^n$ (i.e., axion DM field) and to the classical magnetic field $B_x$, showing relative enhanced sensitivity to low-frequency anomalous axion DM field at the compensation point.
    (b) Frequency-dependent response measured by physically rotating the sensor along the sensitive axis using a custom rotation platform near the compensation point. Deviating from the compensation point reduces the low-frequency response to rotations, while increasing the response to magnetic fields. (c) The compensation-point data from (b), shown separately for clarity. At 0.1 Hz, the equivalent magnetic response to rotation is approximately one order of magnitude larger than that to an applied magnetic field. The error bars denoting one standard deviation are too small to be visually resolved. The response to axion DM field from simulation is also shown and is nearly identical to the response to rotations, due to the (relative to magnetic fields) larger coupling of rotations to nuclear spins compared to electron-based spins.
    (d) As a demonstration of the accuracy of the sensor, a measurement is made with the sensitive axis reoriented to lie parallel to the ground (unlike in the DM search), and a rotation platform is used to rotate the sensitive axis within the horizontal plane. The measured data over one full revolution follows the sinusoidal projection of the Earth's rotation onto the sensitive axis. The error bars denote one standard deviation, but are too small to be visually resolved.}
    \label{fig:calRotation}
\end{figure*}

\begin{figure}[htp!]
    \centering
\includegraphics[width=0.99\linewidth]{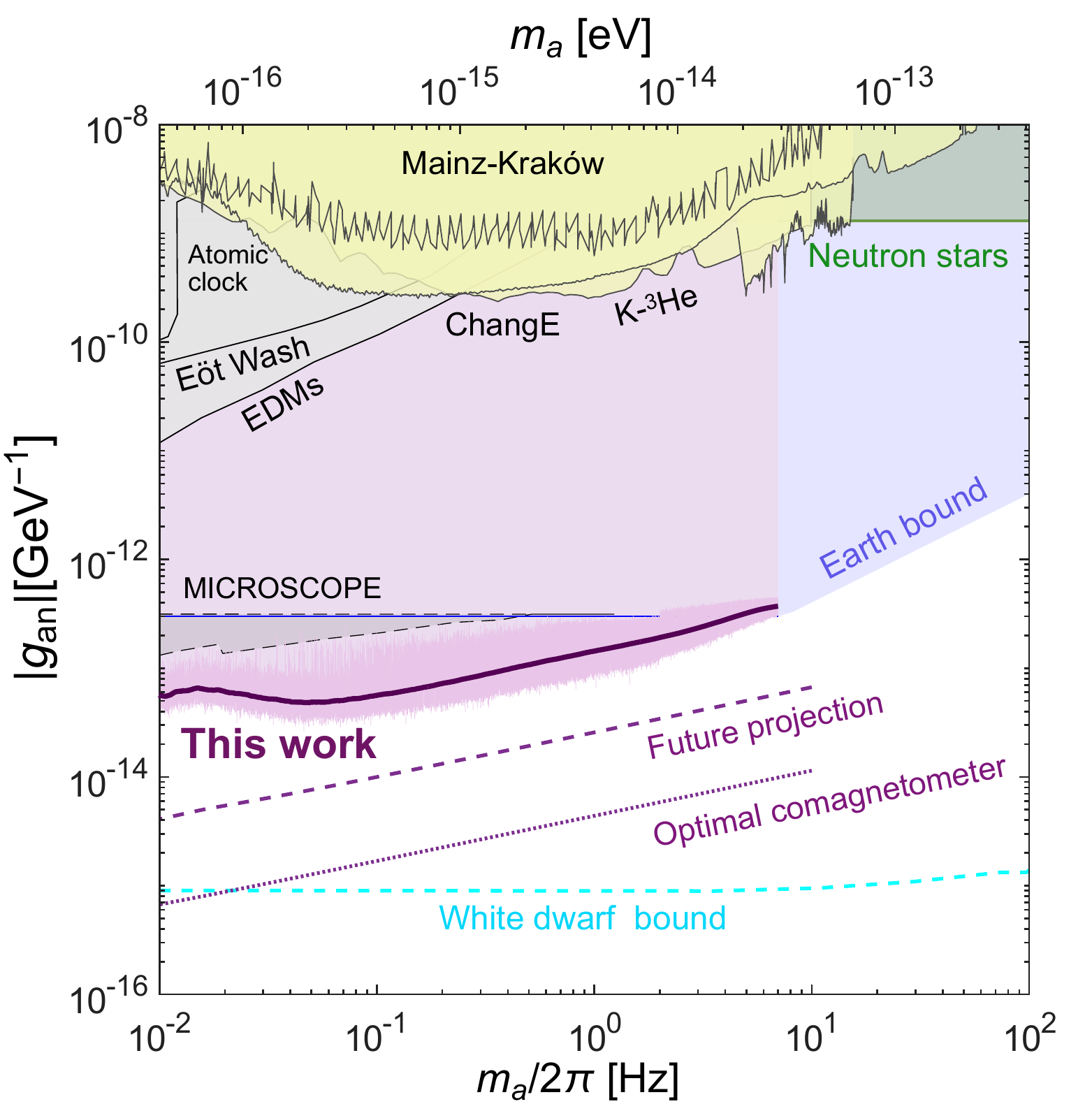}
\caption{ Constraints from the search on the axion-neutron coupling from the data, as well as future projections. The purple solid line show our $95\%$ C.L. exclusion limit of this experiment, with the averaged of the bound plotted in solid dark purple line calculated in a log space at binning resolution of 5\% of the mass.
The purple dashed and dotted lines show the future projections and the sensitivity of the theoretically optimal comagnetometer, respectively. 
By aligning the sensitive axis radially, we maximize the coupling to the Earth-induced axion field gradient, achieving a significant sensitivity enhancement. 
Existing laboratory constraints on axion-neutron coupling, which ignored the enhanced gradient (light-yellow) include comagnetometers~\cite{Lee:2022vvb,Gavilan-Martin:2024nlo,Bloch:2019lcy}, ChangE~\cite{wei2025dark,xu2024constraining}, Mainz-Kraków~\cite{Gavilan-Martin:2024nlo}, with other constraints beyond the bounds of the plot and are not shown~\cite{Bloch:2021vnn,Bloch:2022kjm,Abel:2022vfg,Garcon:2019inh,Wu:2019exd}.
The gray regions are from constraints placed directly on $f_a$ and arise from EDM measurements~\cite{Abel:2017rtm,Roussy:2020ily,JEDI:2022hxa}, atomic clocks~\cite{Madge:2024aot,Zhang:2022ewz} and the MICROSCOPE experiment~\cite{Gue:2025nxq} and are also relevant (see text for further discussion). The bounds on $f_a$ from similar type of experiments~\cite{Schulthess:2022pbp,Roussy:2020ily,Zhang:2022ewz,Fan:2024pxs} are not shown.
The blue shaded region (``Earth Bound'') is excluded to prevent the Earth from sourcing an axion-field new minimum.
Astrophysical bounds from white-dwarfs (cyan dashed)~\cite{Balkin:2022qer} and neutron-star cooling (green shaded)~\cite{Buschmann:2021juv,Springmann:2024mjp} are provided for context, however, these astrophysical bounds rely heavily on complex modeling of stellar equations of state; thus, they suffer from significant systematic uncertainties, emphasizing the importance of the more controlled terrestrial searches. 
Our results surpass previous laboratory limits by 2-3 orders of magnitude and probe the parameter space beyond the Earth-sourced bound. 
In this plot we set $c_n=10$ ($g_{an}= c_n/f_a$); see Fig.~\ref{fig:scalingcn} for other $c_n$ values.}
    \label{fig:finallimit}
\end{figure}

Owing to the different gyromagnetic ratios and coherence times, the responses of alkali-metal and noble-gas spins to external fields vary greatly between them. This dynamical separation enables a special operating condition known as the compensation point~\cite{kornack2005nuclear} at which the optically measured response of the coupled system to magnetic perturbations is strongly suppressed, while the sensitivity to anomalous interactions is preserved, as shown in Fig.~\ref{fig:illu}(a). The noble-gas nuclear spins adiabatically follow external magnetic fluctuations, generating an effective field that cancels the external interference and effectively shields the alkali-metal electron spins from magnetic perturbations.
As the bias field $B_z$ is modified, the sensitivities to axion DM field which couple solely to nuclei and the residual response to magnetic noise, behave differently. Based on the dynamic evolution governing the hybrid spin ensembles~\cite{wei2022constraints,Wei2023}, we simulate the comagnetometer’s response ratio to the anomalous axion DM field $b_y^n$ compared to a classical magnetic field $B_x$ over the frequency range of $10^{-3}-1$ Hz, and over a longitudinal magnetic field range of $\pm 200$ nT around the compensation point ($\delta B_z=0$). The results, shown in Fig.~\ref{fig:calRotation}(a), demonstrate significant suppression of the response to classical magnetic-field perturbations at the compensation point. Thus, near the compensation point, the response to anomalous interactions is significant while the magnetic response reaches a minimum, making it an optimal working point for new physics pseudomagnetic field searches. This operating regime thus allows the probe, which only interacts with the alkali metals, to read out the signal of interest while suppressing the dominant common-mode magnetic background.

A custom single-axis rotation platform is operated in an oscillatory mode, with the angular displacement along the $\hat{y}$ axis, given by $\theta_y=A\sin(2\pi ft)$, yielding an angular velocity drive $\Omega_y=2\pi fA\cos(2\pi ft)$,  where the amplitude $A$ is dynamically adjusted for each frequency to maintain a peak angular velocity within the range of 0.1--0.2~$^\circ$/s. From the measured signal, we obtain the frequency response to rotation, shown as the green circle-marked trace in Fig.~\ref{fig:calRotation}~(b) and (c). For convenience, the corresponding coefficient is converted to an equivalent magnetic response (V/nT) using the nuclear gyromagnetic ratio $\gamma_n$. Under the same conditions, the response to an applied magnetic field along $\hat{x}$, which is the dominant transverse response to magnetic fields, is shown by the yellow square-marked trace. The comagnetometer exhibits a significantly larger response to rotation than to magnetic fields due to strong magnetic-field suppression at the compensation point, which diminishes as the longitudinal field is detuned, in agreement with Fig.~\ref{fig:calRotation}(a). 
Given the agreement between model simulations and experimental data, we use the simulated anomalous response as a reliable calibration to compute the comagnetometer response to DM.
A total of 132 hours of active data taking were recorded.
 We calibrated the system every four hours to monitor the status during data acquisition and ensure the system stability.  

Additionally, we reconfigure the comagnetometer such that the sensitive axis lies in the horizontal plane parallel to the Earth’s surface and use the Earth’s rotation speed as an angular-velocity stimulus. During one full revolution of the apparatus about its vertical axis, the projection of the Earth’s rotation onto the sensitive axis varied sinusoidally; the corresponding measurements (Fig.~\ref{fig:calRotation}(d)) confirms robust sensitivity to this rotation.

\section{Previous limits and model constraints}

For light QCD axion DM, the mass region $f_a\lesssim 10~\rm Hz$ satisfies $kR_{\rm E}<1$, which is one of the requirements to achieve a sizable impact on the axion gradient profile on the surface. The other requirement $q\gtrsim \mathcal{O}(k)$ requires the mass shift $\delta m> k$, which from \eqref{eq:massshift} and the modified mass-decay constant relations gives an upper bound of $\epsilon\lesssim 2\times 10^{-8}$. However, for sufficiently small $\epsilon$ and $\delta mR_{\rm E}\gtrsim 1$, the axion can be driven from perturbations around $a=0$ to ones around $a=\pi f_a$, which is excluded for a variety of reasons and invalidates some of the assumptions in our calculation~\cite{Balkin:2021zfd,Banerjee:2025dlo}. We denote this part of the parameter space as the ``Earth bound'' in Fig.~\ref{fig:finallimit}.
There is a similar astrophysical bound coming from white dwarfs~\cite{Balkin:2022qer}, but this bound relies heavily on complex modeling of the white dwarfs' equation of state, and thus carries far more significant systematic uncertainties compared to
direct laboratory searches. 

In Fig.~\ref{fig:finallimit}, we show the constraints from previous experiments 
from the K--$^3$He comagnetometer~\cite{Lee:2022vvb,Bloch:2019lcy}, Mainz--Krakow comagnetometers~\cite{Gavilan-Martin:2024nlo}, ChangE~\cite{wei2025dark,xu2024constraining}, with NASDUCK~\cite{Bloch:2021vnn,Bloch:2022kjm},  
${}^{199}\mathrm{Hg}$ precession~\cite{Abel:2022vfg}, and CASPEr-ZULF~\cite{Garcon:2019inh,Wu:2019exd} being omitted for the limited range of the frame.
These previous experiments ignored in their analysis the axion-gluon coupling, and were not necessarily aligned with the Earth's radial direction, and hence are only presented here for reference

Given the necessity of the coupling to gluons, this model is also constrained by effects arising from that coupling alone, \textit{e.g.}, electric dipole moment (EDM) measurements~\cite{Schulthess:2022pbp,Abel:2017rtm,Roussy:2020ily,JEDI:2022hxa}, atomic clocks~\cite{Madge:2024aot,Zhang:2022ewz} and constraints from recasting the equivalence principle test of the MICROSCOPE experiment~\cite{Gue:2025nxq}. The recasted limits from MICROSCOPE also took the matter-effect of Earth into consideration, though are weaker than those derived in this work, and have a different dependence on $c_n$. 

\section{Results}

In Fig.~\ref{fig:finallimit},  we show the 95\% C.L. upper limits in purple for coupling $g_{\rm an}$, set as $g_{\rm an}=c_n/f_a$ assuming an axion only coupled to neutrons with the ${}^{21}\rm Ne$ neutron spin fraction $\epsilon_n=0.19$, normalized by the ${}^{21}{\rm Ne}$ spin $3/2$~\cite{Almasi:2018cob,brown2017nuclear}, and the choice $c_n=10$
aligning with Ref.~\cite{Banerjee:2025dlo} (results for larger $c_n$ and related discussion could be found in Fig.~\ref{fig:scalingcn}). 
For the given axion test frequency range, to set a limit on a contiguous segment of axion masses, a total of more than 8 million test are performed. It is also worth noting that due to the isotropy of the leading order signal, there is no significant daily modulation of the signal from Earth's rotation,, despite the long data acquisition time.
This limit on the matter-effect modified axion dark matter profile reaches down to coupling strengths of the order $\mathcal{O}(10^{-13} )~{\rm GeV}^{-1}$, greatly surpassing the sensitivities of experiments which ignored the coupling to gluons, as well as the astrophysical bounds from neutron stars on this coupling~\cite{Buschmann:2021juv,Springmann:2024mjp}. The search for possible 5$\sigma$ signal candidates was performed using similar techniques refs.~\cite{Lee:2022vvb,wei2025dark,xu2024constraining} with no such events found (see Supplemental Material~\cite{sm}).

The near-future projection, as well as the expectations from a more ambitious optimized comagnetometer are plotted as purple dashed and dotted lines in Fig.~\ref{fig:finallimit}, respectively. Our projections assume a pseudo-magnetic field sensitivity $0.1~{\rm fT}/\sqrt{\rm Hz}$ (close to what has already been achieved by ref~\cite{Vasilakis:2008yn} at their quietest frequency), and one year data taking time with 10 correlated sensors. The optimal projection assumes the use of a high-temperature Rotating Wave comagnetometer with $1~\rm aT/\sqrt{\rm Hz}$~\cite{Bloch:2024uqb} sensitivity, 3-year data taking time, and 20 correlated sensors. This optimized system has the ability to probe beyond the WD's estimated limits.  For concreteness and simplicity, the evaluation of the test statistics is based on Asimov dataset~\cite{Cowan:2010js}. Within our regime, ${\bf b}\propto g_{\rm aNN}\partial_ra\propto g_{\rm aNN}^3/c_n^2=c_n/f_a^3$. Therefore, for a fixed $c_n$, further improvements in sensitivity only slowly improve on $g_{\rm aNN}$, though it would also be of interest to probe smaller $c_n$ for a fixed $f_a$ (the WD bound is on $f_a$ alone).

\subsection{The look-elsewhere effects}
To provide a robust limit on all axion masses within the frequency range of interest, we adopt a frequency test spacing of $\Delta f = 1/(2\tau_a)$ with coherence time $\tau_a\sim1/(m_av_{\rm vir}^2)$ and $v_{\rm vir}$ the viral velocity of DM in galaxy ~\cite{Lee:2022vvb}, resulting in over 8 million test points across the searched range. Given the high density of independent trials, the look-elsewhere effect must be accounted for to determine global statistical significances. 

\begin{figure*}[htp!]
  \centering
    \includegraphics[width=0.4\linewidth]{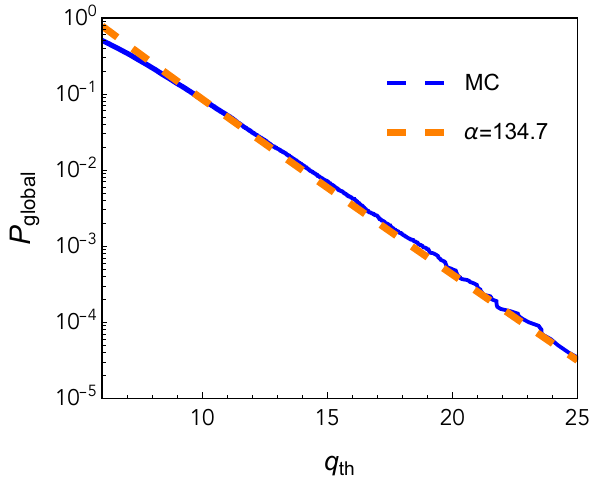}
\includegraphics[width=0.4\linewidth]{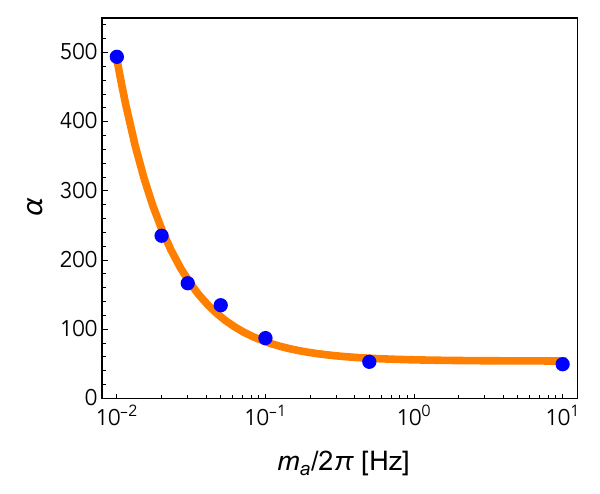}
    \caption{\textit{left panel}: The solid blue line is the global $p$-value $p_{\rm global}$ as a function of the $q_{\rm th}$ obtained from a null Monte Carlo test for axion mass region centered around $0.05~{\rm Hz}$, with the dashed orange line showing its fit. \textit{right panel}: The blue dots represent the Monte Carlo experiments of the corresponding particle frequency. The orange line shows the interpolating function for the  $\alpha(\nu)$ using the blue dots. The notations and Monte Carlo method are adapted from ref.~\cite{Lee:2022vvb}.  }
    \label{fig:lookelse}    
\end{figure*}

Due to the intrinsic axion linewidth and the finite frequency resolution of the Fourier-transformed experimental data, constraints on neighboring masses are highly correlated. In Fig.~\ref{fig:lookelse}, we determine the effective number of independent tests, $N$, using a Monte Carlo methodology consistent with Ref.~\cite{Lee:2022vvb}. A detailed discussion of the trial factor, the impact of data acquisition time, and the numerical integration of $N$ is provided in the Supplemental Material~\cite{sm}. Incorporating the LEE, the global $5\sigma$ detection threshold is established at likelihood ratio $\text{LLR}_{\text{significance}} = 48.51$ with detailed definitions in the Supplemental Material~\cite{sm}. As shown in Fig.~\ref{fig:Sensitivity}, no candidates reached this global significance threshold in our experiment.

\subsection{The Results for Other Choices of $c_n$}\label{sec:otherchoices}

In main text, we parameterize the axion-neutron strength via $g_{\rm an}=c_n/f_a$, with $c_n=10$ chosen to present the final results. In this section we show and discuss other values of $c_n$.

\begin{figure}[htp!]
    \centering
    \includegraphics[width=0.9\linewidth]{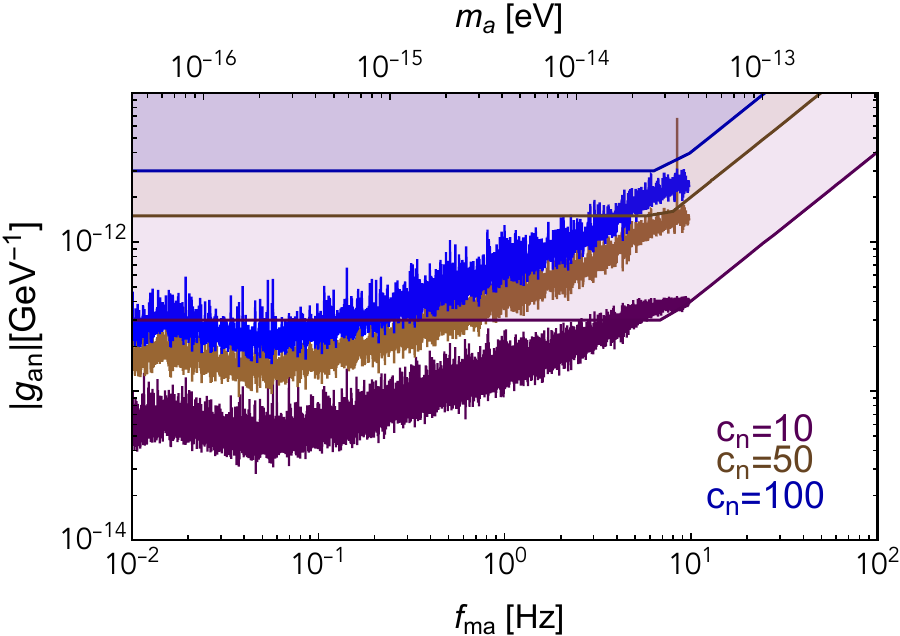}
    \caption{The final limits and the corresponding Earth bounds shown in purple ($c_n=10$), brown ($c_n=50$), blue ($c_n=100$). The Earth bounds directly depends on decay constant $f_a$, thus scales linearly with $c_n$, while the experimental constraints scale as $c_n^{-3/2}$ in this regime.}
    \label{fig:scalingcn}
\end{figure}

Fig.~\ref{fig:scalingcn} shows the limits on the coupling $g_{\rm an}$ for other choices of $c_n=50/100$. Generally for larger $c_n$ our limits are less stringent.  This could be understood from the simple qualitative behavior Eq.~\ref{eq:radialGradientEarth}. The observed signal strength is proportional to $g_{\rm an}\partial_r a$, and could be rewritten as $ g_{\rm an}^3 /c_n^2$. one can see clearly that for a fixed signal strength, increasing the $c_n$ would leads to a weaker constrain on the coupling strength $g_{\rm an}$. On the other hand, since the Earth and white dwarfs bounds solely depend on the decay constant $f_a$, increasing the $c_n$ would results in less constrained Earth and white dwarf bounds on the $g_{\rm an}$ parameter space.
Smaller choices of $c_n$, quickly encounter the Earth-bound region, or can excite resonances for the higher masses, requiring a more dedicated work beyond the scope of this work. We note that we focused on $c_n=10<4\pi$, as it is not immediately obvious that larger values are possible from the model-building perspective. While generic ALPs often have their shift symmetric coupling $g_{\rm aNN}$ far larger than their couplings to gluons, it is not obvious that the same can be done when looking at light-QCD axion models without breaking unitarity.

\section{Conclusion}

We report experimental results on the search for light QCD axion DM coupling to neutrons using alkali-noble-gas hybrid spin ensembles. Benefiting from the enhanced axion gradient field on Earth, we set a 95\% C.L. upper limits on the axion--neutron coupling $g_{\rm aNN}\sim \mathcal{O}(10^{-13})\,\rm GeV^{-1}$ in the Compton frequency band $0.01$--$7$~Hz, which represents a substantially improved limit compared with existing constraints on standard axion dark matter model. We find no candidate signal events, demonstrating the discovery-grade quality of our sensors. As our results lie below the ``Earth Bound'', they are squarely in a parameter domain where our calculation is valid and robust, demonstrating that our sensitivity is capable of probing viable new-physics space rather than merely approaching a future target.
Importantly, these laboratory-based constraints provide a direct and controlled exploration of Earth matter-effected axion dark matter that is complementary to astrophysical arguments, remaining robust even in model variations where stellar bounds may be weakened or evaded by other potential new physics contributions~\cite{Masso:2005ym,DeRocco:2020xdt}.
Furthermore, the well-defined orientation of the sensitive axis provides access to the spatial structure of the effective field induced by massive bodies, allowing the experiment to exploit geometric modulation of the signal while suppressing common-mode backgrounds.

Unlike the naive expectation from the Standard Halo Model, the Earth-matter-effected axion framework exhibits a specific radial dependence in the effective pseudo-magnetic field, which is maximized at the surface and diminishes rapidly with increasing altitude. Taking axion Compton frequency $m_a/2\pi=10~{\rm Hz}$ and decay constant $f_a=10^{14} ~\rm {GeV}$ as an example, at a space station 400 km above the Earth surface,  the signal strength decreases to its 90\% compared with surface value, while on the orbit of the James Webb space telescope, the strength could further decrease to 1\%. 
This unique gradient feature serves as a definitive validation test for this particular model. To probe this phenomenology, we plan to deploy our comagnetometer systems on the China Space Station (CSS) in Low Earth Orbit (LEO), forming a strategic baseline with terrestrial detectors.
The resulting space-ground network will allow for a direct differential measurement of the axion field gradient profile.
Such a configuration can leverage network-correlation techniques to veto local systematics, thereby significantly extending the experimental sensitivity and enabling a robust discovery potential, rather than merely excluding regions of parameter space or yielding ambiguous “hints.”

\begin{acknowledgments}
We thank Sebastian A. R Ellis, Jordan Gue and Jia Liu for useful discussions. This work is supported by  NSFC Grants No.62473022, by the Innovation Program for Quantum Science and Technology under Grant 2021ZD0300401.
\end{acknowledgments}

\bibliography{Bibli}

\clearpage        
\appendix
\clearpage        
\setcounter{page}{1}
\begin{widetext}
\setcounter{figure}{0}
\setcounter{table}{0}
\setcounter{equation}{0}

\renewcommand{\thefigure}{S\arabic{figure}}
\renewcommand{\thetable}{S\arabic{table}}
\renewcommand{\theequation}{S\arabic{equation}}

\section*{Supplemental Material}

\subsection{The Stochastic Treatment of Axion DM Field with Matter Effects}\label{sec:stochastic}
Much like in the usual Standard Halo Model~\cite{Lee:2022vvb,Foster:2017hbq,Bloch:2021vnn}, the matter-effected axion DM solution is a linear superposition of different velocity modes.
For the frequency range of interest where $kR_{\rm E}\ll 1$, the  spherically symmetric $\ell=0$ contributions dominate the matter-effected axion DM gradient field, making the modulation from Earth's rotation negligible, and thus simplifying the analysis.

We model the standard vacuum axion field far away from Earth without matter effects as the superposition of various velocity modes.
The form of the non-relativistic axion dark matter field within the Galactic halo away from Earth and other condensed matter can be expressed as~\cite{Lee:2022vvb,Foster:2017hbq}:
\begin{align}
 a(\bm r,t)\bigg|_{\bm r \gg R} 
\approx& \sum_{\Omega_\mathbf{\bm p}} \frac{\sqrt{\rho_\textrm{DM} f(\mathbf{v}) (\Delta v)^3}}{m_a} \, \alpha_\mathbf{v} \times \cos\left(m_a \left(1 + \frac{1}{2}v^2\right)t-m_a \mathbf{v} \cdot \bm r + \Phi_\mathbf{v}\right)\,,
\end{align}

where $\rho_{\rm DM}=0.4~\text{GeV}/c^2\text{cm}^3$~\cite{de_Salas_2021} represents the local DM density, $f(\bf{v})$ denotes the DM velocity distribution function under the assumption of the Standard Halo Model~\cite{Lee_2013}, 
$\alpha_\mathbf{v}$ follows standard Rayleigh distribution as a random variable, and $\Phi_\mathbf{p}$ represents a random phase uniformly distributed within $\left[0,2\pi\right]$. The summation is taken over all the infinitesimal volume elements in momentum space, denoted as $\Omega_\mathbf{p}$. 
We choose this formula as the boundary condition at infinity. By identifying
\begin{align}
\frac{\sqrt{\rho_\textrm{DM} f(\mathbf{v}) (\Delta v)^3}}{m_a} \alpha_\mathbf{v} \leftrightarrow 2a_0(\bm v)
\end{align} 
in  ~\eqref{eq:AsympAxionDMSol} and according to energy conservation and the superposition principle for homogeneous solutions of differential equations, 
we could directly write down the $\ell=0$ contribution,
\begin{align}
    \partial_r a(R_{\rm E},t)|_{\ell =0}&=\sum_{\Omega_{\mathbf{v}}} \alpha_{\mathbf{v}}\frac{\sqrt{\rho_{\rm DM} f(\mathbf{v}) (\Delta v)^3}}{m_a} \times \nonumber\\
    &\frac{ \big(q R_{\rm E} \cos (q R_{\rm E})-\sin (q R_{\rm E})\big) }{2R_{\rm E}^2 \left(k^2 \sin ^2(q R_{\rm E})+q^2 \cos ^2(q R_{\rm E})\right)}\big[(k+q) \cos (k R_{\rm E}-q
   R_{\rm E}-t \omega -\phi(\bm v) )+(q-k) \cos (k R_{\rm E}+q R_{\rm E}-t \omega -\phi(\bm v)
   )\big]\,,
   \label{eq:stochastic}
\end{align}
where we used $k=m_a v$, $q=\sqrt{\dm +m_a^2v^2}$ and $\omega=m_a(1+v^2/2)$. To simplify our notations, we define $\chi(v)\equiv \big(q R \cos (q R)-\sin (q R)\big) /\big[2R^2 \left(k^2 \sin ^2(q R)+q^2 \cos ^2(q R)\right)\big]$ hereafter.
For $kR_{\rm E}\ll q R_{\rm E} \lesssim \mathcal{O}(1)$, the second line in ~\eqref{eq:stochastic} asymptotically approach to $-\frac{1}{3}q^2R_{\rm E}\cos(q R_{\rm E})\cos(\omega t+\phi(\bm v))$,  and around a factor of $1/3 q^2R_{\rm E}^2\cos(qR_{\rm E})/(kR_{\rm E})$ improvement compared with standard axion dark matter model, with $qR_{\rm E}\lesssim1$ this leads to around $1/kR_{\rm E}\times (f_a^c)^2/f_a^2$ factor enhancement as stated previously.  For $qR_{\rm E}\ll1$, this asymptotes to $qR_{\rm E}$, at which point one must add the contribution from the $\ell=1$ mode to see no enhancement or suppression is present for objects which are too small. 

As mentioned before, within our region of interest, the $\ell =0$ contribution to the gradient field dominates on the surface, and is sufficient for the analysis in this work. A Remarkable feature of this radial gradient field is that it is homogeneous, and so as mentioned before, unlike previous works  (e.g. Refs.~\cite{wei2025dark,xu2024constraining,Lee:2022vvb,Bloch:2022kjm}), no substantial daily modulation is present in the signal.
Taking our sensitive axis to be pointing upwards, we have the measured effective magnetic field $B_{\rm eff}=2g_{\rm aNN}/\gamma_n \bm \nabla a\cdot \hat{n}=2\gn/\gamma_n \partial_ra$. For a measured time data series ${\beta_N}= \beta_0 ...\beta_{N-1}$ with sampling interval $\Delta T$ and total time $T$, we define the real and imaginary part of the amplitude in frequency domain:
\begin{equation}
    A_k = \frac{2}{N}{\bf Re}[\tilde{\beta}_k] \,,\, B_k = -\frac{2}{N}{\bf Im}[\tilde{\beta}_k]\,,
    \label{eq:Ak_Bk_def}
\end{equation}

 with $\tilde{\beta}_k$ as the discrete Fourier transform of obtained data time series $\beta_{n}$,
\begin{equation}
    \tilde{\beta}_k = \sum_{n=0}^{N-1} \beta_{n} \exp^{-{\rm i}\omega_k n \Delta t} \,,\, \omega_k \equiv \frac{2{\rm \pi} k}{N\Delta t}\,.
    \label{eq:beta_fft_def}
\end{equation}

After Fourier transforming, we can utilize the following formula,
\begin{align}
    \sum_{n=0}^{N-1} \Delta t \cos(\omega n \Delta t +\phi)e^{-i \omega_k n \Delta t}\approx e^{i[(\omega-\omega_k)T/2+\phi]}{\rm sinc}[(\omega-\omega_k)T/2]T/2,
\end{align}
with the approximation holding for $(\omega-\omega_k)\Delta T \approx 0$, which is indeed the case.  Following similar derivation steps in Refs.~\cite{Lee:2022vvb,Bloch:2021vnn} and carefully treating the statistic properties of $\alpha_{\bm v}$ and $\phi(\bm v)$ we find the axion dark matter effective magnetic field has a mean value of 0 and its covariance matrix in the frequency domain is given by:
\begin{align}
\nonumber
\sigma(A_k, A_r) &= \rho_\textrm{DM} \left(\frac{g_\textrm{aNN}}{m_aT\mu_\textrm{Ne}}\right)^2  \int d\bm v \, f(\bm v) \chi(v)^2\left[E_{k}(\bm v)E_{r}(\bm v) + F_{k}(\bm v)F_{r}(\bm v)\right],
\label{eq:covAA_def}
\end{align}
\begin{align}
\nonumber
\sigma(A_k, B_r) &= \rho_\textrm{DM} \left(\frac{g_\textrm{aNN}}{m_aT\mu_\textrm{Ne}}\right)^2 \int d\bm v \, f(\bm v) \chi(\bm v)^2\left[E_{k}(\bm v)F_{r}(\bm v) - F_{k}(\bm v)E_{r}(\bm v)\right],
\end{align}
\begin{align}
\nonumber
\sigma(B_k, A_r) &= -\sigma(A_k, B_r)=\sigma(A_r, B_k),\\
\sigma(B_k, B_r) &= \sigma(A_k, A_r).\nonumber
\end{align}

With functions $E_k(\bm v)$, $E_k(\bm v)$ defined as 
\begin{align}
\displaystyle 
& E_{k}(\bm v) = \frac{T}{2}{\rm sinc}[(\omega-\omega_k)T/2]\left(
 (q + k)\,\cos\left[ \frac{T\,(\omega - \omega_k)}{2} + (q - k)\,R \right]
 +
 (q - k)\,\cos\left[ \frac{T\,(\omega - \omega_k)}{2} - (q + k)\,R \right]
\right),\nonumber\\
& F_{k}(\bm v) = -\frac{T}{2}{\rm sinc}[(\omega-\omega_k)T/2]\left(
 (q + k)\,\sin\left[ \frac{T\,(\omega - \omega_k)}{2} + (q - k)\,R \right]
 +
 (q - k)\,\sin\left[ \frac{T\,(\omega - \omega_k)}{2} - (q + k)\,R \right]
\right)\,.
\end{align}

Because of the isotropy of this solution, the angular integration becomes simpler than the more usual case where Earth's rotation becomes important.

\subsection{Likelihood Analysis and Sensitivity Derivation}
\label{sec:llr}
We conduct an analysis using the Log-Likelihood Ratio (LLR) method based on the response power spectrum $(V^2)$. The calibration function $g_\text{cali}(f)$ here is the transfer function that converts anomalous magnetic field strength in Tesla to the apparatus measured data in Voltage at a given frequency $f$. We obtain this function by measuring the system response to injected transverse magnetic fields of known amplitudes and varying their frequency. From the fit parameters, we can then derive the response to the axion DM field, see details in the \textit{Calibration of the frequency response} Section. This aforementioned calibration procedure is conducted before each measurement.

Since the signal is sampled from a multivariate Gaussian distribution, and the background is modeled as white noise with variance $\Sigma_{\rm n}$ within a small frequency window around the target axion frequency, we can construct the likelihood function as follows: 
        \begin{align}
            L(\textbf{d}|g_{\rm aNN}, \Sigma_{\rm n})=\frac{1}{\sqrt{(2\pi)^{2N}\mathrm{det}(\Sigma)}} \mathrm{exp}\left(-\frac{1}{2}\textbf{d}^T\Sigma^{-1}\textbf{d}\right).
        \end{align}
with the vector ${\bf d} = \left\{ A_k, B_k\right\} $ and the covariance matrix $\Sigma=\Sigma_{\rm A}(g_{\rm aNN})+\Sigma_{\rm n}\cdot \mathbbm{1}$.  The signal covariance matrix is defined as $\Sigma_{\rm A}(\gn)(f_k,f_r)=g_\text{cali}(f_k)g_\text{cali}(f_r)\begin{pmatrix}
 \sigma(A_k,A_r)& \sigma(A_k,B_r)\\
 \sigma(B_k,A_r)&\sigma(B_k,B_r)
 \end{pmatrix}$.
To set our limit, we use the LLR defined below~\cite{PDG}
                \begin{align}
                    {\rm LLR}(g_{\rm aNN}) =
                    \begin{cases}
                        -2\log\left[\frac{L(g_{\rm aNN},\tilde{\Sigma}_{\rm n})}{L(\hat{g}, \hat{\Sigma}_{\rm n})}\right],&\hat{g}\leq  g_{\rm aNN} \\
                        0, &\hat{g}> g_{\rm aNN}
                    \end{cases}
                \end{align}
where we treat $\Sigma_{\rm n}$ as nuisance parameters to be marginalized, to maximizes the likelihood $L$ for a fixed nucleus coupling $g_{\rm aNN}$ in each evaluation. $\hat{\Sigma}_{\rm n}$ and $\hat{g}$ are unconditional estimators that maximize likelihood $L$ for a given $\bf d$. In order to establish an upper limit on the coupling, we set ${\rm LLR}=0$ if $\hat{g} > g_{\rm aNN}$. The $ {\rm LLR}$ can be shown to asymptotically follow a half-$\chi^2$ distribution with the cumulative distribution function~\cite{Cowan:2010js}
\begin{align}
    \Phi( {\rm LLR}(g_{\rm aNN})\le y)=\frac{1}{2}\left[1+\text{erf}\left(\sqrt{\frac{y}{2}}\right)\right] ,
    \label{eq:cdf}
\end{align}
thus the $95\%$ CL upper limits are determined by finding the value of $g_{\rm aNN}$ where ${\rm LLR}(g_{\rm aNN})=2.7055$.
The effective coupling between axion and the nucleus $N$ is given by $g_{\rm aNN} = \xi_{\rm n} g_{\rm aNN} + \xi_{\rm p} g_{\rm app}$, where $\xi_{\rm n}^{\rm Ne} = 0.19$ and $\xi_{\rm p}^{\rm Ne} = 0.013$ are the spin-polarization fractions for neutrons and protons in ${}^{21}{\rm Ne}$ normalized by its spin $3/2$~\cite{Almasi:2018cob,brown2017nuclear,xu2025prl}. 

Due to the velocity dispersion, the axion dark matter typically has a frequency linewidth roughly $\Delta f_a=1/\tau_a$ with the coherence time $\tau_a\sim1/(m_av_{\rm vir}^2)$~\cite{Lee:2022vvb,Bloch:2021vnn,Centers:2019dyn}. We choose the axion frequency test spacing with $\Delta f=1/2\Delta f_a$ to provide a robust continuous limit on the axion mass parameter space, ending up with over eight million test points.

In addition to a constraints, we also search for a possible positive detection. To find the possible signals that are unlikely arise due to statistical fluctuations of the background, we define the test statistics to quantify the significance of a best-fit signal compared with background only model, which is the method that has been adopted in Refs.~\cite{wei2025dark,xu2024constraining}:
 \begin{align}
                    {\rm LLR}_{\rm significance} =
                    \begin{cases}
                        -2\log\left[\frac{L(0,\tilde{\Sigma}_{\rm n})}{L(\hat{g}, \hat{\Sigma}_{\rm n})}\right],&\hat{g}\leq  0, \\
                        0, &\hat{g}> 0.
                    \end{cases}
                \end{align}
This test statistics also follows the half-$\chi^2$ distribution assuming the signal-free model, and the $p$-value $p$ is simply $p=1-\Phi(\sqrt{ {\rm LLR}_{\rm significance}})$, leading to the $5\sigma$ detection threshold being ${\rm LLR}_{\rm significance}^{5\sigma}=22.7$ without considering the look-elsewhere effects.
The above statement is valid for a single point test. However, since our experiment searches for a wide range of axion masses, with millions of independent test points, the look-elsewhere effect plays an important role. For $N$ independent tests, the global $p$-value $p_{\rm global}$ is effectively $p_{\rm global}\approx N p$ for sufficiently small $p$. Since the frequency data points overlap for nearby axion test frequencies, one typically need to rely on Monte Carlo method to determine the effective number $N$ of the independent test. In this paper we adopt the same Monte Carlo method as in Ref.~\cite{Lee:2022vvb}, with the same notations. To briefly summarize, assuming the axion masses $m_a^{(i)}=m_a^{(0)}(1+\alpha v_0^2)^i$ are independent, at  asmall frequency region $[f_{\rm min},~f_{\rm max}]$ centered around a  specific point, we make a large numbers of Monte Carlo tests with the same axion test frequency grid and same effective time as the experiment data,  then using the maximum ${\rm LLR}_{\rm significance}$ (defined as $q_{\rm th}$) of each test to formulate a practical cumulative distribution function (CDF) and fit it with theoretical predictions to get $\alpha(v)$, see fig.~\ref{fig:lookelse}. The reason that the value of $\alpha(v)$ is much smaller compared with that in ref.~\cite{Lee:2022vvb} is the much short data acquisition time, the frequency interval $1/T_{\rm eff}$ is much larger, leading to more overlapping of the axion linewidth, corresponding to smaller independent numbers ($N\propto 1/\alpha$). The final number of $N$ is obtained via integration of the logarithmic space $N=\int_{\rm log ~min(f)}^{\rm log~max(f)}1/(\alpha(v)v_{\rm vir}^2) {\rm d ~log}(v)$, with  ${\rm log ~min(f)}/{\rm log ~max(f)}$ being the minimum and maximum frequencies of the entire searched range. Combined with the Monte Carlo results we get the $5~\sigma$ detection threshold ${\rm LLR}_{\rm significance}^{5\sigma,{\rm look-elsewhere}}=48.51$ taking into account of the look-elsewhere effect.
Thanks to the noise control of the experiment on the frequency range of the interest, we find no candidates that reach the global $5~ \sigma$ threshold in this experiment (c.f. the data of Fig.~\ref{fig:Sensitivity}).

\subsection{Dynamics of Alkali-Noble-Gas Hybrid Spin Ensembles}
\label{sec:dynamics}
The Hamiltonians for the alkali and noble-gas are
\begin{align}\label{eqn-31}
\left\{ \begin{array}{l}
\mathcal{H}^{\rm{e}} =A_g\textbf{I}\cdot\textbf{S}+g_e \mu_B\textbf{S}\cdot\textbf{B}-\frac{\mu_I}{I}\textbf{I}\cdot\textbf{B}+g_e\mu_B\textbf{S}\cdot{\boldsymbol{b}}^{\rm{e}}\,,\\
\nonumber\\
\mathcal{H}^{\rm{n}} =-\frac{\mu_K}{K}\textbf{K}\cdot\left(\textbf{B}+{\boldsymbol{b}}^{\rm{n}}\right)\,,
\end{array}\right.
\end{align}
where $\textbf{I}$ and $\textbf{S}$ are the spin operators of alkali nucleus and electron, respectively. $\textbf{K}$ is nuclear spin of noble gas. $\textbf{B}$ is the magnetic field. $A_g$ represents the ground state hyperfine coupling constant. $g_e$ is the electron spin g-factor. $\mu_B$ is the Bohr magneton. $\mu_I$ is the nuclear dipole moment of alkali atom, and $\mu_K$ is the nuclear dipole moment of noble gas. $\boldsymbol{b}^{\rm{e}}$ denotes the anomalous fields (e.g., axion DM field) that couples to the alkali spins, and ${\boldsymbol{b}}^{\rm{n}}$ represents that couples to the noble-gas spins. Due to the interaction between the two, the magnetic terms actually partially overlap between the two hamiltonians.

Due to the rapid spin exchange between potassium and rubidium (with a spin exchange rate on the order of $10^5 \, \text{s}^{-1}$), the electron spins dynamics of the hybrid ensembles are essentially consistent and can be equivalently described using ${\bf{P}}^{\bf{e}}$~\cite{Chen2016Spin,Wei2021Accurate}. ${\bf{P}}^{\rm{n}}$ represents the polarization of nuclear spins. The evolution of the hybrid spin ensembles can be described by the Bloch equations
\begin{equation}\label{eqn-32}
\left\{ \begin{array}{l}
\displaystyle\frac{\partial{\bf{P}}^{\rm{e}}}{\partial t}= \frac{{{\gamma _e}}}{Q}\left({\bf{B}} +\bf{L}+ \lambda M_0^n{{\bf{P}}^{\rm{n}}} + {{\boldsymbol{b}}^{\rm{e}}}\right) \times {{\bf{P}}^{\rm{e}}} + \frac{{{R_p}{{\bf{S}}_{\bf{p}}} + {R_m}{{\bf{S}}_{\bf{m}}} + R_{\rm{se}}^{\rm{ne}}{{\bf{P}}^{\bf{n}}}}}{Q} - \frac{{\{ R_2^e,R_2^e,R_1^e\} }}{Q}{{\bf{P}}^{\rm{e}}}\,,\\[8pt]
\displaystyle\frac{\partial{\bf{P}}^{\rm{n}}}{\partial t} = {\gamma _n}\left({\bf{B}} + \lambda M_0^e{{\bf{P}}^{\rm{e}}}+{{\boldsymbol{b}}^{\rm{n}}}\right)\times {{\bf{P}}^{\rm{n}}} + R_{\rm{se}}^{\rm{en}}{{\bf{P}}^{\rm{e}}} - \{ R_2^n,R_2^n,R_1^n\} {{\bf{P}}^{\rm{n}}}\,,
\end{array} \right.
\end{equation}
where $\gamma_e$ is the gyromagnetic ratio of electrons spin. $Q$ is the slowing down factor. ${\bf{L}}$ is the light shift. $R_p$ and $R_m$  are the mean pumping rates of unpolarized atoms of the ground state by the pump light and the probe light. $\lambda=8\pi\kappa_0/3$ and $\kappa_0$ is the Fermi-contact enhancement factor. $M^{\rm{e/n}}_0$ represents the magnetizations of electron or nuclear spins corresponding to full spin polarizations. ${{\bf{S}}_{\bf{p}}}$ and ${{\bf{S}}_{\bf{m}}}$ are the photon spin of pump light and probe light. $R_{\rm{se}}^{\rm{ne}}$ and $R_{\rm{se}}^{\rm{en}}$ are the spin-exchange rates experienced by alkali and noble-gas, respectively. $R_1^{\rm{e/n}}$ and $R_2^{\rm{e/n}}$ are the longitudinal and transverse relaxation rates of alkali electron spins or noble-gas nuclear spins.

\subsubsection{Response to magnetic field}
The compensation point mentioned in the main text is defined as the condition  where the longitudinal magnetic field $B_c=-B_0^n-B_0^e=-\lambda M_0^nP_z^n-\lambda M_0^e P_z^e$. When the transverse perturbations of the magnetic field are sufficiently small, the longitudinal polarizations of the alkali-metal electron spins $P_z^e$ and noble-gas nuclear spins $P_z^n$  remain approximately constant. By defining the $P_ \bot ^e = P_x^e + iP_y^e$ and $P_ \bot ^n = P_x^n + iP_y^n$, the transverse dynamical evolution of the coupled spin ensembles in the presence of the magnetic field $B_ \bot  = B_x + iB_y$ can then be written as 

\begin{equation}
\frac{\partial }{{\partial t}}\left[ \begin{array}{l}
P_ \bot ^e\\
P_ \bot ^n
\end{array} \right] = \left[{\begin{array}{*{20}{c}}
{ - i\frac{{{\gamma _e}}}{Q}B_0^e - \frac{{R_2^e}}{Q}}&{ - i\frac{{{\gamma _e}}}{Q}\lambda M_0^nP_z^e}\\
{ - i{\gamma _n}\lambda M_0^eP_z^n}&{ - i{\gamma _n}B_0^n - R_2^n}
\end{array}}\right]\left[ \begin{array}{l}
P_ \bot ^e\\
P_ \bot ^n
\end{array} \right] + \left[ \begin{array}{l}
 - i\frac{{{\gamma _e}}}{Q}{B_ \bot }P_z^e\\
 - i{\gamma _n}{B_ \bot }P_z^n
\end{array} \right]
\end{equation}

As for the low-frequency magnetic field ${B_ \bot }\left( t \right) = \left( {{B_x} + i{B_y}} \right)\cos \left( {\omega t} \right) = \frac{1}{2}{B_ \bot }\left( {{{\rm{e}}^{i\omega t}} + {{\rm{e}}^{ - i\omega t}}} \right)$, we can solve for the positive- and negative-frequency components separately. As the positive-frequency components, we have 
$\left[ \begin{array}{l}
{P_ \bot ^{e}}^{\rm{pos}}\left( t \right)\\
{P_ \bot ^{n}}^{\rm{pos}}\left( t \right)
\end{array} \right] = \left[ \begin{array}{l}
{P_ \bot ^{e}}^{\rm{pos}}{{\rm{e}}^{i\omega t}}\\
{P_ \bot ^{n}}^{\rm{pos}}{{\rm{e}}^{i\omega t}}
\end{array} \right]$, and
$\frac{\partial }{{\partial t}}\left[ \begin{array}{l}
{P_ \bot ^{e}}^{\rm{pos}}\left( t \right)\\
{P_ \bot ^{n}}^{\rm{pos}}\left( t \right)
\end{array} \right] = i\omega \left[ \begin{array}{l}
{P_ \bot ^{e}}^{\rm{pos}}{{\rm{e}}^{i\omega t}}\\
{P_ \bot ^{n}}^{\rm{pos}}{{\rm{e}}^{i\omega t}}
\end{array} \right]$. Thus, the steady-state response is

\begin{equation}
\begin{aligned}
   \left[ \begin{array}{l}
{P_ \bot ^{e}}^{\rm{pos}}\\
{P_ \bot ^{n}}^{\rm{pos}}
\end{array} \right] 
= &{\left( {i\omega  - \left[ {\begin{array}{*{20}{c}}
{ - i\frac{{{\gamma _e}}}{Q}B_0^e - \frac{{R_2^e}}{Q}}&{ - i\frac{{{\gamma _e}}}{Q}\lambda M_0^nP_z^e}\\
{ - i{\gamma _n}\lambda M_0^eP_z^n}&{ - i{\gamma _n}B_0^n - R_2^n}
\end{array}} \right]} \right)^{ - 1}}\left[ \begin{array}{l}
 - i\frac{{{\gamma _e}}}{Q}{B_ \bot }P_z^e\\
 - i{\gamma _n}{B_ \bot }P_z^n
\end{array} \right] \\
= &- \frac{{\left[ \begin{array}{l}
\left( {\omega  - iR_2^n} \right)\frac{{{\gamma _e}}}{Q}{B_ \bot }P_z^e\\
\left( {\omega  - i\frac{{R_2^e}}{Q}} \right){\gamma _n}{B_ \bot }P_z^n
\end{array} \right]}}{{\omega \left( {\omega  + \frac{{{\gamma _e}}}{Q}B_0^e + {\gamma _n}B_0^n} \right) - \frac{{R_2^e}}{Q}R_2^n - i\left( {\left( {\omega  + \frac{{{\gamma _e}}}{Q}B_0^e} \right)R_2^n + \frac{{R_2^e}}{Q}\left( {\omega  + {\gamma _n}B_0^n} \right)} \right)}}\,.
\end{aligned}
\end{equation}

Similarly, the expression for the negative-frequency component is given by
\begin{equation}
\left[ \begin{array}{l}
{P_ \bot ^{e}}^{\rm{neg}}\\
{P_ \bot ^{n}}^{\rm{neg}}
\end{array} \right] =  - \frac{{\left[ \begin{array}{l}
\left( {\omega  + iR_2^n} \right)\frac{{{\gamma _e}}}{Q}{B_ \bot }P_z^e\\
\left( {\omega  + i\frac{{R_2^e}}{Q}} \right){\gamma _n}{B_ \bot }P_z^n
\end{array} \right]}}{{\omega \left( { - \omega  + \frac{{{\gamma _e}}}{Q}B_0^e + {\gamma _n}B_0^n} \right) + \frac{{R_2^e}}{Q}R_2^n + i\left( {\left( { - \omega  + \frac{{{\gamma _e}}}{Q}B_0^e} \right)R_2^n + \frac{{R_2^e}}{Q}\left( { - \omega  + {\gamma _n}B_0^n} \right)} \right)}}\,.
\end{equation}

Thus, the response of the low-frequency magnetic field is $\left[ \begin{array}{l}
{P_ \bot ^{e}}(t)\\
{P_ \bot ^{n}}(t)
\end{array} \right] 
=\frac{1}{2}\left[ \begin{array}{l}
{P_ \bot ^{e}}^{\rm{pos}}\\
{P_ \bot ^{n}}^{\rm{pos}}
\end{array} \right] {\rm{e}}^{i\omega t}+
\frac{1}{2}\left[ \begin{array}{l}
{P_ \bot ^{e}}^{\rm{neg}}\\
{P_ \bot ^{n}}^{\rm{neg}}
\end{array} \right] {\rm{e}}^{-i\omega t} $.

When the transverse magnetic field oscillates with frequency $\omega$ ($\omega \ll \gamma_n B_0^n$) and amplitudes $B_x$ and $B_y$ , the effective magnetic field produced by nuclear spins has components along the $\hat{x}$ and $\hat{y}$ given by
\begin{align}
\begin{aligned}
B_x^n(\omega) =\lambda M_0^n P_x^n(\omega)=  & - {B_x}\cos \left( {\omega t} \right)- \frac{{{\gamma _e}B_0^{e}R_2^n}}{{R_2^{e}{\gamma _n}B_0^n}}{B_x}\cos \left( {\omega t} \right) - \frac{{\omega {\gamma _e}B_0^{e}}}{{R_2^{e}{\gamma _n}B_0^n}}{B_x}\sin \left( {\omega t} \right) \\
& + \frac{{R_2^n{B_y}\cos \left( {\omega t} \right)}}{{{\gamma _n}B_0^n}} - \frac{{\omega {B_y}\sin \left( {\omega t} \right)}}{{{\gamma _n}B_0^n}}\,,\\
B_y^n (\omega)=\lambda M_0^n P_y^n(\omega)=  &- {B_y}\cos \left( {\omega t} \right)- \frac{{R_2^n{B_x}\cos \left( {\omega t} \right)}}{{{\gamma _n}B_0^n}} + \frac{{\omega {B_x}\sin \left( {\omega t} \right)}}{{{\gamma _n}B_0^n}}\\
& - \frac{{{\gamma _e}B_0^{e}R_2^n}}{{R_2^{e}{\gamma _n}B_0^n}}{B_y}\cos \left( {\omega t} \right) - \frac{{\omega {\gamma _e}B_0^{e}}}{{R_2^{e}{\gamma _n}B_0^n}}{B_y}\sin \left( {\omega t} \right)\,.\label{eq:BnBxBycos}
\end{aligned}
\end{align}

The first term on the right side of both equations represents the self-compensation of the nuclear spin, where the effective magnetic field of the nuclear spin counteracts the external magnetic field, while the remaining terms account for interference contributions.  It can be seen that reducing the transverse nuclear spins relaxation rate $R_2^n$ is beneficial for minimizing extra interference contributions. The nuclear spins transverse relaxation time is predominantly limited by the effective magnetic field gradient arising from the gradient of the electron spins polarization. To suppress this effective field gradient, it is advantageous to polarize a small fraction of optically thin K atoms and then transfer the polarization to a large number of Rb atoms via spin exchange, thereby improving the polarization homogeneity, as shown in Fig.~\ref{fig:HP}.

\begin{figure*}[!ht]
    \centering
    \includegraphics[width=0.7\linewidth]{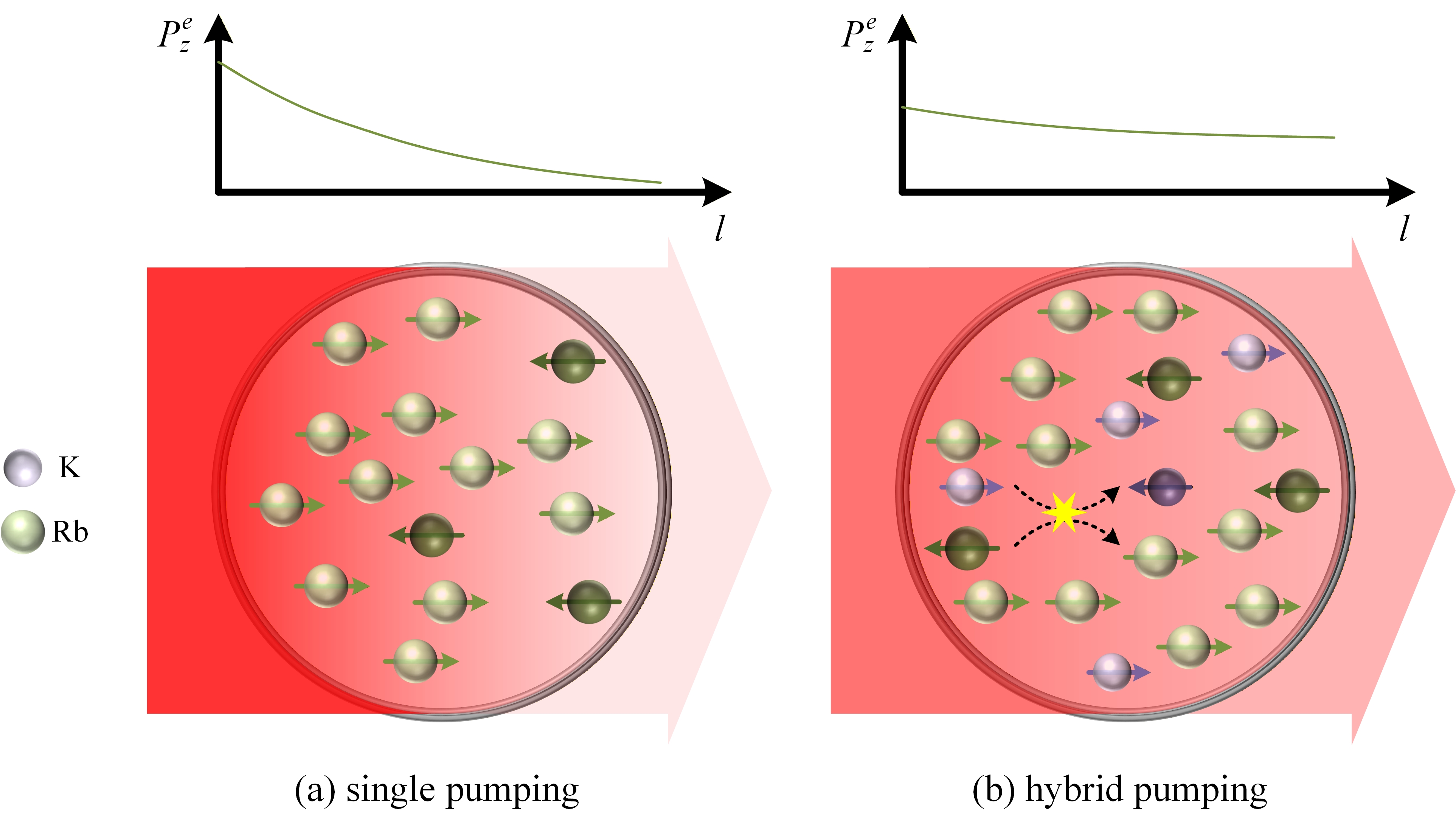}
    \caption{Comparison between single and hybrid Alkali pumping. (a) When only a single alkali species is present, the high atomic number density leads to strong absorption of the pump light along its propagation path. Although the polarization in the center of the cell reaches the optimal value of $\approx 0.5$, the polarization is high near the entrance and low near the exit, resulting in a pronounced polarization gradient. The associated gradient of the electron effective magnetic field reduces the nuclear-spin transverse relaxation time. (b) When a small amount of optically thin K is uniformly polarized by the pump light, rapid spin-exchange collisions subsequently produce a homogeneous polarization of Rb, thereby effectively reducing the electron polarization gradient.}
    \label{fig:HP}
\end{figure*}

Thus, the comagnetometer output signal (with the nuclear spins relaxation neglected), which is the electron polarization along the $\hat{x}$ is approximately 

\begin{align}
\begin{aligned}
P_x^e \approx & {B_x}\frac{{{\gamma _e}P_z^e}}{{R_2^e}} \left( {\frac{\omega }{{{\gamma _n}B_0^n}} - \frac{{2R_2^e{\omega ^3}}}{{R_2^e\left({\gamma _n}B_0^n \right )^3}}} \right)\sin{\omega t}  - B_x\frac{{{\gamma _e}P_z^e}}{{R_2^e}} \left( \frac{{Q{\omega ^2}}}{{R_2^e{\gamma _n}B_0^n}} + \frac{{{\gamma _e}B_0^e{\omega ^2}}}{{R_2^e{{\left( {{\gamma _n}B_0^n} \right)}^2}}} \right)  \cos \omega t\\
& + B_x\frac{{{\gamma _e}P_z^e}}{{R_2^e}}\frac{{2{\omega ^4}Q}}{{R_2^e  {{{\left( {{\gamma _n}B_0^n} \right)}^3}} }} \cos \omega t \\
&- {B_y}\frac{{{\gamma _e}P_z^e}}{{R_2^e}} \left( {  \frac{{Q{\omega ^3}}}{{{{\left( {{\gamma _n}B_0^n} \right)}^2}R_2^e}} + \frac{{2{\omega ^3}{\gamma _e}B_0^e}}{{R_2^e{{\left( {{\gamma _n}B_0^n} \right)}^3}}}} \right)\sin \omega t - {B_y}\frac{{{\gamma _e}P_z^e}}{{R_2^e}} \frac{{{\omega ^2}}}{{{{\left( {{\gamma _n}B_0^n} \right)}^2}}}\cos \omega t\,.\label{eq:PexBxBycos}
\end{aligned}
\end{align}
It can be seen that at zero frequency the response to the magnetic field vanishes. As the frequency increases, this suppression gradually weakens.

\subsubsection{Response to axion DM field}

Similarly, as for the response to $b_y^n \cos(\omega t)=b_y^n({\rm{e}}^{i\omega t}+{\rm{e}}^{-i\omega t})$, the transverse polarization of coupled spin ensemble are
\begin{equation}
\left[ \begin{array}{l}
P_ \bot ^e\\
P_ \bot ^n
\end{array} \right] =  - \frac{1}{2}\left\{
\begin{aligned}
&\frac{{\left[ \begin{array}{l}
 - i\frac{{{\gamma _e}}}{Q}{\gamma _n}B_0^nP_z^eb_y^n\\
\left( {i\omega  + i\frac{{{\gamma _e}}}{Q}B_0^e + \frac{{R_2^e}}{Q}} \right){\gamma _n}b_y^nP_z^n
\end{array} \right]}}{{\omega \left( {\omega  + \frac{{{\gamma _e}}}{Q}B_0^e + {\gamma _n}B_0^n} \right) - \frac{{R_2^e}}{Q}R_2^n - i\left( {\left( {\omega  + \frac{{{\gamma _e}}}{Q}B_0^e} \right)R_2^n + \frac{{R_2^e}}{Q}\left( {\omega  + {\gamma _n}B_0^n} \right)} \right)}}{{\rm{e}}^{i\omega t}}\\
&+\frac{{\left[ \begin{array}{l}
i\frac{{{\gamma _e}}}{Q}{\gamma _n}B_0^nP_z^eb_y^n\\
\left( {i\omega  - i\frac{{{\gamma _e}}}{Q}B_0^e - \frac{{R_2^e}}{Q}} \right){\gamma _n}b_y^nP_z^n
\end{array} \right]}}{{\omega \left( { - \omega  + \frac{{{\gamma _e}}}{Q}B_0^e + {\gamma _n}B_0^n} \right) + \frac{{R_2^e}}{Q}R_2^n + i\left( {\left( { - \omega  + \frac{{{\gamma _e}}}{Q}B_0^e} \right)R_2^n + \frac{{R_2^e}}{Q}\left( { - \omega  + {\gamma _n}B_0^n} \right)} \right)}}{{\rm{e}}^{ - i\omega t}}\,.
\end{aligned}
\right\}
\end{equation}

At the condition of $\omega$ ($\omega \ll \gamma_n B_0^n$), we have
\begin{equation}
\begin{aligned}
B_x^n = \lambda M_0^nP_x^n = & - \frac{\omega }{{{\gamma _n}B_0^n}}b_y^n\sin \left( {\omega t} \right) + \frac{{\omega {\gamma _e}^2B{{_0^e}^2}}}{{R{{_2^e}^2}{\gamma _n}B_0^n}}b_y^n\sin \left( {\omega t} \right) + \frac{{\omega \gamma_e QB_0^e}}{{R{{_2^e}^2}}}b_y^n\sin \left( {\omega t} \right)\\
&+ \frac{{R_2^n}}{{{\gamma _n}B_0^n}}b_y^n\cos \left( {\omega t} \right) + \frac{{{\gamma _e}^2B{{_0^e}^2}R_2^n}}{{R{{_2^e}^2}{\gamma _n}B_0^n}}b_y^n\cos \left( {\omega t} \right) + \frac{{B_0^e{\gamma _e}}}{{R_2^e}}b_y^n\cos \left( {\omega t} \right)\\
B_y^n = \lambda M_0^nP_y^n =&  - b_y^n\cos \left( {\omega t} \right) - \frac{{2\omega {\gamma _e}^2B{{_0^e}^2}R_2^n}}{{R{{_2^e}^2}{{\left( {{\gamma _n}B_0^n} \right)}^2}}}b_y^n\sin \left( {\omega t} \right)\\
&- \frac{{2\omega {\gamma _e}B_0^e}}{{R_2^e{\gamma _n}B_0^n}}b_y^n\sin \left( {\omega t} \right) - \frac{{2\omega R_2^n}}{{{{\left( {{\gamma _n}B_0^n} \right)}^2}}}b_y^n\sin \left( {\omega t} \right) + \frac{{2\omega {\gamma _e}QB_0^eR_2^n}}{{R{{_2^e}^2}{\gamma _n}B_0^n}}b_y^n\sin \left( {\omega t} \right)
\end{aligned}
\end{equation}

In this equation, only the first-order term in frequency is retained. The first term of $B_y^n$ on the right-hand side represents an effective magnetic field generated by the nuclear spins, which is equal in magnitude and aligned in direction with the anomalous field, and can be sensed by the electron spins. It should be noted that, for a classical magnetic field, the electron spins experience both the field itself and an effective field generated by the nuclear spins that is equal in magnitude but opposite in direction, thereby realizing self-compensation. In contrast, for an anomalous field that couples only to the nuclear spins, the electron spins do not sense the field directly; their response arises solely from the transfer of the effective field induced by the nuclear spins. It also can be seen in the response of transverse electron spins polarization $P_x^e$ to the $b_y^n$ as

\begin{equation}
P_x^e =  - {\gamma _e}P_z^e\left[ {\left( {\frac{{2\omega R_2^n}}{{R_2^e{{\left( {{\gamma _n}B_0^n} \right)}^2}}} + \frac{{\omega {\gamma _e}B_0^e}}{{R{{_2^e}^2}{\gamma _n}B_0^n}} + \frac{{Q\omega }}{{R{{_2^e}^2}}}} \right)b_y^n\sin \left( {\omega t} \right) + \left( {\frac{{{\gamma _e}B_0^eR_2^n}}{{R{{_2^e}^2}{\gamma _n}B_0^n}} + \frac{1}{{R_2^e}}} \right)b_y^n\cos \left( {\omega t} \right)} \right]\,.\label{eq:bynRep}
\end{equation}

\subsubsection{Self-compensation}
For a deviation ($\delta B_z=B_z - B_c$) from the compensation point, the  steady-state of alkali electron spin polarization to DC input along $\hat{x}$ is
\begin{equation}\label{eqn-33}
P_x^e = \frac{{{\gamma _e}P_z^e}}{{R_2^e}}\left\{ {b_y^e - \left({1+\frac{\delta B_z}{\lambda M_0^n P_z^n}}\right) b_y^n-\frac{\delta B_z}{\lambda M_0^n P_z^n} B_y+\frac{\gamma_e}{R_2^e}{\delta B_z}\left[-b_x^e+\left({1+\frac{\delta B_z}{\lambda M_0^n P_z^n}}\right) b_x^n + \frac{\delta B_z}{\lambda M_0^n P_z^n} B_x \right] } \right\}\,.
\end{equation}

At the compensation point, the electron spins transverse polarization is 
\begin{equation}
P_x^e = \frac{{{\gamma _e}P_z^e}}{{R_2^e}}( {b_y^e - b_y^n } )\,.\label{SCPexSteady}
\end{equation}

From this expression, it is clear that perturbations from external magnetic fields are compensated, leaving only the response to the anomalous fields. This is consistent with the result obtained from Eq.~\ref{eq:bynRep} when evaluated at zero frequency. Based on the above analysis of the responses to both the ordinary magnetic field and the anomalous field, the principle of nuclear spin self-compensation can be understood as follows: the effective magnetic field generated by the nuclear spins cancels the external magnetic perturbation with equal magnitude and opposite direction, so that the electron spins remain unaffected by the disturbance. For an anomalous field that couples only to the nuclear spins, the electron spins serve as a highly sensitive in-situ magnetometer for the nuclear effective magnetic field , thereby enabling its measurement, as shown in Fig.~\ref{fig:SCPri}.

\begin{figure*}[!ht]
    \centering
    \includegraphics[width=0.6\linewidth]{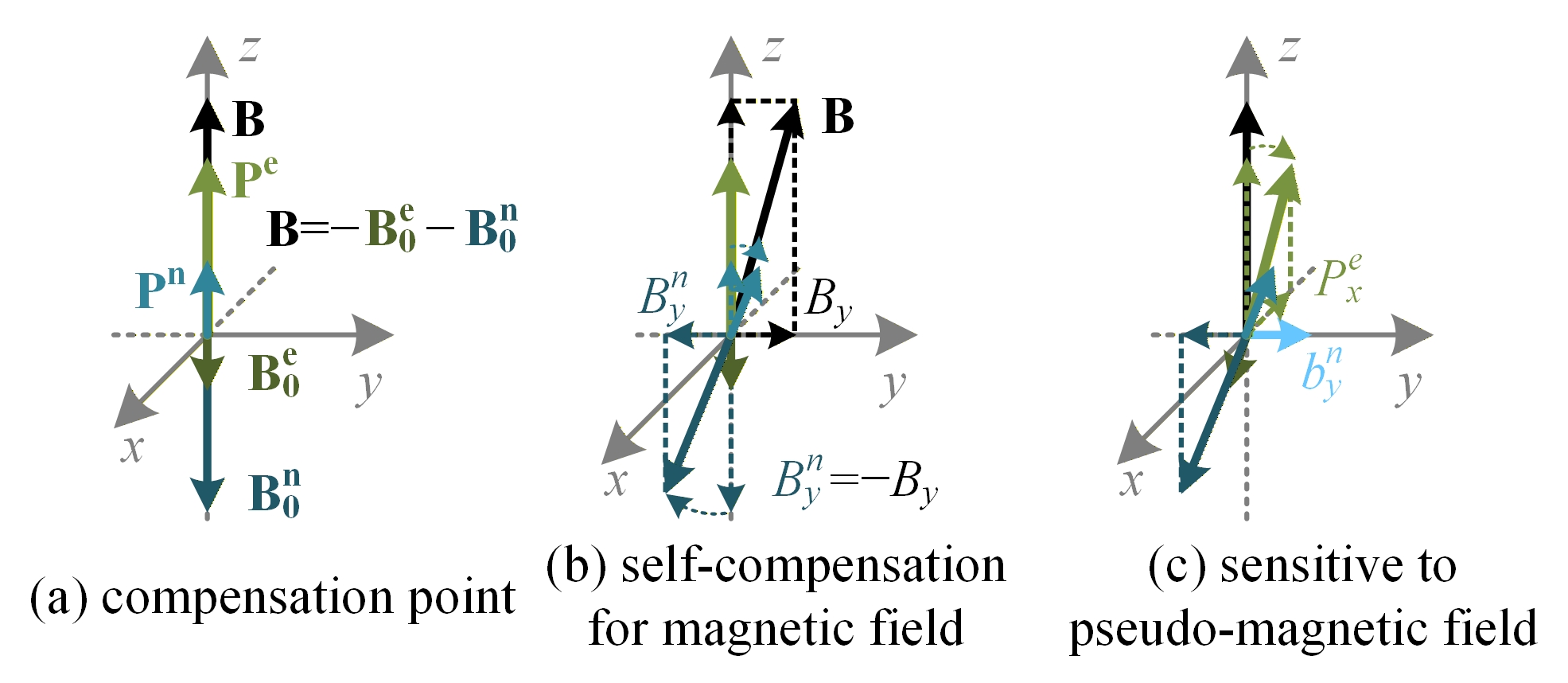}
    \caption{Principle of nuclear self-compensation. (a) Circularly polarized pump light polarizes the electron spins, which subsequently polarize the nuclear spins in the same direction via spin-exchange interactions. The effective magnetic fields generated by both spin ensembles are opposite to their polarization direction. The compensation point is reached when the applied longitudinal magnetic field is equal in magnitude and opposite in direction to the combined effective magnetic fields. (b) When a low-frequency transverse magnetic perturbation is present, the effective magnetic field generated by the nuclear spins develops a transverse component that is equal in magnitude and opposite in direction, thereby suppressing the total magnetic field experienced by the electron spins; this constitutes nuclear-spin self-compensation.  (c) When an anomalous field that couples only to the nuclear spins (e.g., the axion DM field) is present, the electron spins cannot directly sense the anomalous field itself. However, the Fermi-contact enhancement enables the electron spins to act as a highly sensitive in-situ magnetometer that detects the polarization changes induced in the nuclear spins.}
    \label{fig:SCPri}
\end{figure*}

\subsubsection{Calibration of the frequency response}
\label{sec:calibration}

Over the full frequency range, the responses to the magnetic field or the anomalous field is difficult to express in a simplified form. Under small-input assumption, the state-space representation provides an accurate characterization of their frequency-domain behavior. With transverse magnetic fields and $y$-axis anomalous fields coupled with nuclear spins as inputs, the state equation is
\begin{equation}
{\bf{\dot P}} = {\bf{AP}} + {\bf{BU}}\,,
\end{equation}
where ${\bf{P}} = {\left[ {\begin{array}{*{20}{c}}
{P_x^e}&{P_y^e}&{P_x^n}&{P_y^n}
\end{array}} \right]^{\rm{T}}}$ represents the state vector, and ${\bf{U}} = {\left[ {\begin{array}{*{20}{c}}
{{B_x}}&{{B_y}}&{{b_y^n}}
\end{array}} \right]^{\rm{T}}}$ is the input vector. The coefficient matrix $\bf{A}$ is 
\begin{equation}
{\bf{A}} = \left[ {\begin{array}{*{20}{c}}
{ - \frac{{R_2^e}}{Q}}&{ - {\omega _e}}&{\frac{{R_{se}^{ne}}}{Q}}&{{\omega _{en}}}\\
{{\omega _e}}&{ - \frac{{R_2^e}}{Q}}&{ - {\omega _{en}}}&{\frac{{R_{se}^{ne}}}{Q}}\\
{R_{se}^{en}}&{{\omega _{ne}}}&{ - R_2^n}&{ - {\omega _n}}\\
{ - {\omega _{ne}}}&{R_{se}^{en}}&{{\omega _n}}&{ - R_2^n}
\end{array}} \right],
\end{equation}
and  the input matrix $\bf{B}$ is 
\begin{equation}
{\bf{B}} = \left[ {\begin{array}{*{20}{c}}
0&{\frac{{{\gamma _e}}}{Q}P_z^e}&0\\
{ - \frac{{{\gamma _e}}}{Q}P_z^e}&0&0\\
0&{{\gamma _n}P_z^n}&{ \gamma_n P_z^n}\\
{ - {\gamma _n}P_z^n}&0&0
\end{array}} \right],
\end{equation}
with ${\omega _e} = \frac{{{\gamma _e}}}{Q}\left( {{B_z} + {L_z} + \lambda M_0^nP_z^n} \right)$ , ${\omega _{en}} = \frac{{{\gamma _e}}}{Q}\lambda M_0^nP_z^e$ , ${\omega _{ne}} = {\gamma _n}\lambda M_0^eP_z^n$ , and ${\omega _n} = {\gamma _n}\left( {{B_z} + \lambda M_0^eP_z^e} \right)$.

Based on the state equation, we likewise simulate the response ratio between the anomalous field and the $B_y$ magnetic field, as shown in Fig.~\ref{fig:Bybcom}. Fig.~\ref{fig:Bybcom}(a) shows the response ratio evaluated over $\pm 200$ nT range about the compensation point (by tuning the longitudinal magnetic field) for frequencies spanning $10^{-3}$ to 1~Hz. For clarity, Fig.~\ref{fig:Bybcom}(b) presents the corresponding curves at five representative frequencies. The results demonstrate that the compensation point provides excellent suppression of classical magnetic-field perturbations, with an even stronger suppression for $B_y$.

\begin{figure}[!htbp]
    \centering
    \includegraphics[width=0.8\linewidth]{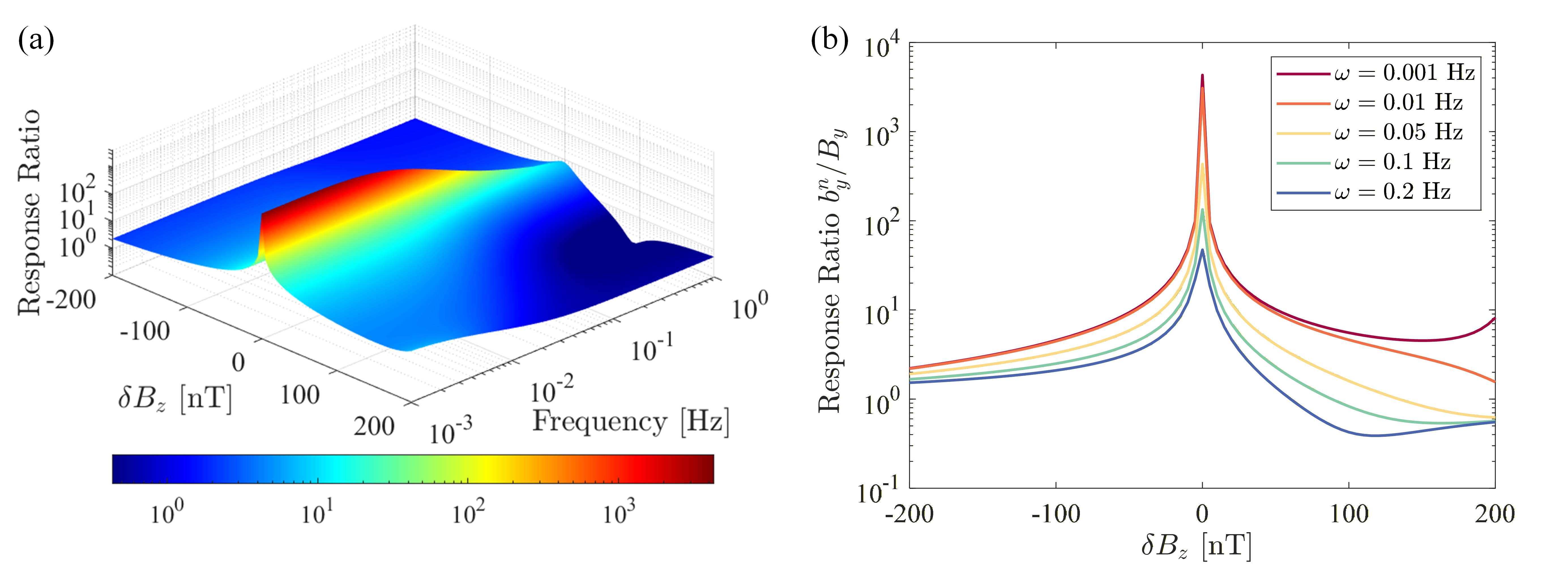}
    \caption{Simulated responses to the anomalous field $b_y^n$ and classical magnetic field $B_y$ based on the dynamical model. (a) The response ratio within $\pm 200$ nT of the compensation point for frequencies from $10^{-3}$ to 1~Hz, with five representative frequency traces shown in (b).}
    \label{fig:Bybcom}
\end{figure}

\subsection{The Detailed Experimental Setup}

As shown in Fig.~\ref{fig:ExSe}, the experimental setup comprises a spherical cell with a $12$\,mm diameter, which contains a small amount of potassium, a droplet of naturally abundant rubidium, along with $^{21}$Ne of $3$\,amg and $0.06$\,amg N$_2$ as quenching gas. The cell is heated to 468\,K by AC electric heater to reduced low-frequency magnetic noise. A four-layer $\mu$-metal shield with an inner Ferrite shield, along with a triaxial coil, provides active and passive magnetic shielding to suppress ambient remanence and magnetic noise. A circular polarized resonant pump light tuned to the D1 line of potassium (770.108\,nm) realizes the polarization of electron spins along $\hat{z}$, and in turn the spin exchange optical pumping of nuclear spins. The linear polarized far-detuned probe light (795.501 nm), modulated by PEM of 50\,kHz propagates along $\hat{x}$ and reads out the polarization of electron spins. The modulated optical rotation is demodulated by the lock-in amplifier at which the sampling rate was 899.5\,Hz, presenting the $P_x^e$.

\begin{figure}[!ht]
    \centering
    \includegraphics[width=0.5\linewidth]{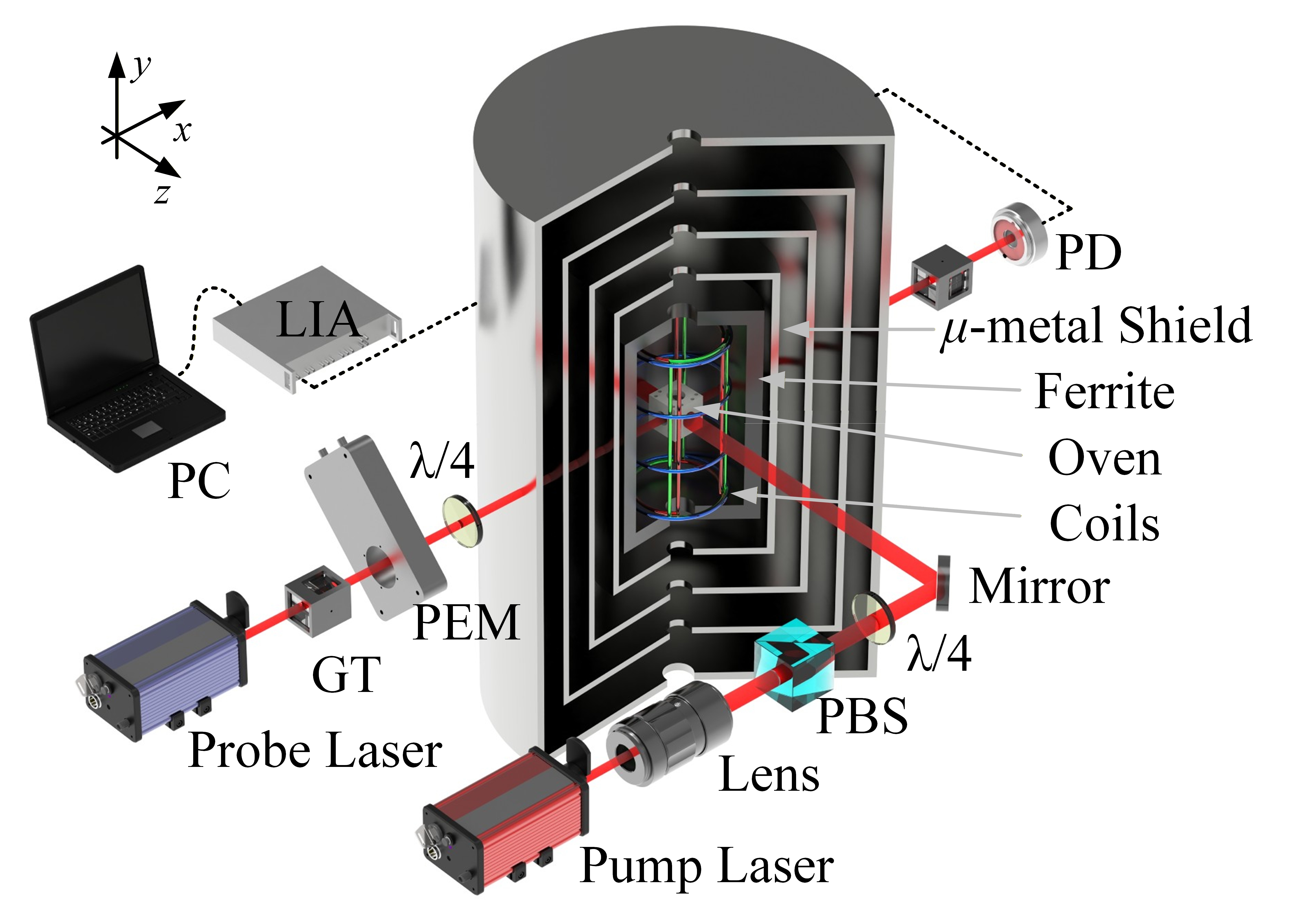}
    \caption{Basic operation of the experimental setup. The sphere cell contains K, Rb, $3$ amg $^{21}$Ne and $0.06$ amg N$_2$. A circularly polarized pump light polarizes the hybrid spin ensembles along $\hat{z}$. A probe light is used to measure the $\hat{x}$ projection of electron polarization, which is modulated by PEM and demodulated by lock-in amplifier. The ceramic oven is equipped with double-layer winding heating wires, in which the current is generated by an AC electric heater. Magnetic shields consist of a four-layer $\mu$-metal shield and a Ferrite inner layer shield the Earth's or laboratory magnetic field. The triaxial coil are used to ensure further compensation of remanence and manipulate that  experienced by spins. PC: personal computer, LIA: lock-in amplifier, PBS: polarization beam splitter, GT: Glan-Taylor polarizer, PEM: photoelastic modulator, PD: photodetector.}
    \label{fig:ExSe}
\end{figure}

\subsection{Sensitivity}

Prior to data acquisition, the comagnetometer is tuned to its optimal sensitivity through a combination of non-orthogonality suppression and systematic parameter optimization. The bias field is then adjusted to locate the compensation point, at which the ratio between anomalous field and magnetic field responses is maximized. Data are subsequently collected continuously for a total duration of 132 h at this operating point. To ensure long-term stability, the working point of the comagnetometer is recalibrated every 4 h throughout the measurement. The acquired time-domain data (in volt) are analyzed in the frequency domain to obtain the amplitude spectral density (in V/$\sqrt{\rm{Hz}}$). Dividing by the calibrated frequency-dependent response factor (in V/fT) then yields the sensitivity, based on which we can calculate the energy resolution by $\gamma_nB\hbar$ as shown in Fig.~\ref{fig:Sensitivity}.

\begin{figure*}[!ht]
    \centering
    \includegraphics[ width=0.5\textwidth]{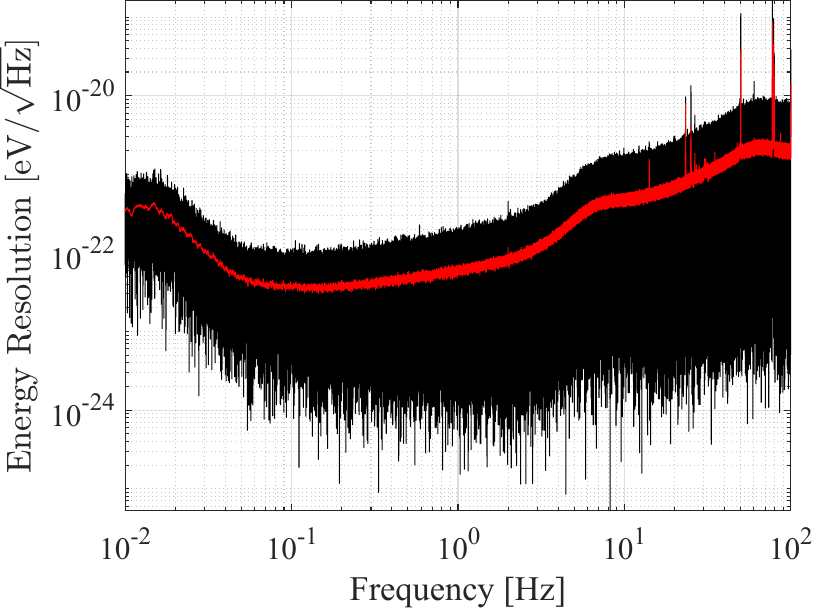}
    \caption{Sensitivity to nuclear spin coupled anomalous fields along $\hat{y}$. We obtain the sensitivity by spectral analysis of the time-domain signal and dividing it by the calibrated response factor at each frequency. The energy resolution, given by $\gamma_n B\hbar$, is $3.8\times 10^{-23} \mathrm{eV/\sqrt{Hz}}$ at 0.1\,Hz, corresponding to a sensitivity of 2.7 fT/$\sqrt{\rm Hz}$.
    }
    \label{fig:Sensitivity}
\end{figure*}

At present, the dominant noise sources in our comagnetometer include polarization noise, magnetic noise, vibration noise, thermal noise, and probe noise. \textit{Polarization noise:} Polarization noise arises from optical pumping polarization effects: fluctuations in the power, frequency, or polarization state of the circularly polarized pump beam modulate the pumping rate, thereby perturbing all components of the spin polarization. In addition, the residual circular-polarization component in the probe light can still couple its fluctuations into the measurement, introducing additional noise. \textit{Magnetic noise:} Although our multilayer $\mu$-metal and ferrite shielding provides an attenuation factor of $10^6$, the hysteresis-loss noise of the shield  is approximately 3 fT/$\sqrt{\rm{Hz}}$ at 1 Hz , following an $1/\sqrt{f}$ dependence as a function of frequency. In addition, the non-magnetic heating assembly inside the shield can introduce weak magnetic noise due to structural design imperfections or machining tolerances. While nuclear-spin self-compensation offers a degree of active magnetic-noise suppression, further improvements in sensitivity will require an even more stable magnetic environment. \textit{Vibration noise:} Alkali–noble-gas comagnetometers are inherently sensitive to rotational signals. Human activities, mechanical vibrations, acoustic disturbances, and microseismic ground motion can introduce angular-velocity components that manifest as spurious anomalies in the measured signal. Implementing a hybrid active–passive vibration-isolation system would help suppress low-frequency vibrations and thereby improve the low-frequency sensitivity. \textit{Thermal noise:} Thermal noise mainly manifests in two aspects: convection-induced fluctuations and temperature variations within the cell. Thermal convection in the apparatus can lead to beam-pointing instability along the optical path; placing the core components (e.g., the cell and optical paths) in a vacuum environment would mitigate this disturbance. In addition, fluctuations in the heating power can change the atomic number density in the cell, which in turn directly affects key parameters such as the relaxation rates of both spin species and ultimately the measured signal output. Thermal convection noise primarily manifests in the low-frequency range, below 0.01 Hz up to 1 Hz. \textit{Probe  noise:} We employ a photoelastic modulator to shift the signal to a higher frequency, thereby mitigating low-frequency disturbances. However, the modulation-demodulation chain itself can still suffer from limited low-frequency stability and introduce residual interference. Moreover, as the sensitivity is further improved, quantum-noise limits (such as photon shot noise and spin-projection noise) become increasingly prominent. Methods based on squeezing and entanglement therefore merit future implementation.

\end{widetext}

\end{document}